\documentclass[fleqn,usenatbib]{mnras}

\usepackage{float}

\usepackage{newtxtext,newtxmath}

\usepackage{xcolor}
\usepackage{anyfontsize}

\usepackage[LGR, T1]{fontenc}

\usepackage{pifont}

\usepackage{hyperref}
\hypersetup{colorlinks=true, citecolor=blue, linkcolor=blue}

\DeclareTextAccentDefault{\accdasia}{LGR}
\DeclareTextAccentDefault{\acctonos}{LGR}

\DeclareRobustCommand{\VAN}[3]{#2}
\let\VANthebibliography\thebibliography
\def\thebibliography{\DeclareRobustCommand{\VAN}[3]{##3}\VANthebibliography}

\newcommand\textlcsc[1]{\textsc{\MakeLowercase{#1}}}

\usepackage{graphicx}	
\usepackage{amsmath}	
\title[PopIII: Confirmation of HeII at z=10.6]{The search for Population III: Confirmation of a HeII emitter with no metal lines at z=10.6}
\author[Maiolino et al.]{Roberto Maiolino,$^{1,2,3}$ \thanks{E-mail: \href{mailto:rm665@cam.ac.uk}{rm665@cam.ac.uk}}
Hannah \"{U}bler,$^{4}$
Michele Perna,$^{5}$
Joris Witstok,$^{6,7}$
Gareth C. Jones,$^{1,2}$
\newauthor
Pablo G. Pérez-González,$^{5}$
Kimihiko Nakajima,$^{8,9,10}$
Elka Rusta,$^{11,12}$
Stefania Salvadori,$^{11,12}$
\newauthor
Sandro Tacchella,$^{1,2}$
Piero Madau,$^{13,14}$
James A.\@ A.\@ Trussler,$^{15}$
Francesco D’Eugenio,$^{1,2}$
Xihan Ji,$^{1,2}$
\newauthor
Jan Scholtz,$^{1,2}$
Stefano Carniani,$^{16}$
Yuki Isobe,$^{1,2,17}$
Harley Katz,$^{18,19}$
Santiago Arribas,$^{5}$
\newauthor
William M. Baker,$^{20}$
Torsten B\"oker,$^{21}$
Volker Bromm,$^{22,23,24}$
Andrew J.\ Bunker,$^{25}$
Stephane Charlot,$^{26}$
\newauthor
Jacopo Chevallard,$^{25}$
Giovanni Cresci,$^{12}$
Mirko Curti,$^{27}$
Emma Curtis-Lake,$^{28}$
Daniel Eisenstein,$^{15}$
\newauthor
Eiichi Egami,$^{29}$
Andrea Ferrara,$^{16}$
Luca Graziani,$^{30,31}$
Kevin Hainline,$^{29}$
Jakob M. Helton,$^{32}$
\newauthor
Lucy R. Ivey,$^{1,2}$
Benjamin D. Johnson,$^{15}$
Ignas Juod\v{z}balis,$^{1,2}$
Maria Koller,$^{1,2}$
Nimisha Kumari,$^{33}$
\newauthor
Nicolas Laporte,$^{34}$
Alessandro Marconi,$^{11,12}$
Giovanni Mazzolari,$^{4}$
Eleonora Parlanti,$^{16}$
\newauthor
Robert Pascalau,$^{1,2}$
Laura Pentericci,$^{35}$
Pierluigi Rinaldi,$^{36}$
Brant Robertson,$^{14}$
\newauthor
Bruno Rodríguez Del Pino,$^{5}$
Raffaella Schneider,$^{30}$
Alessandra Venditti,$^{22,24}$
Giacomo Venturi,$^{16}$
\newauthor
Christopher N. A. Willmer,$^{29}$
Callum Witten,$^{37}$
Sandra Zamora$^{16}$
}
\begin{document}
\label{firstpage}
\pagerange{\pageref{firstpage}--\pageref{lastpage}}
\maketitle
%
\begin{abstract}
We report the confirmation of a HeII$\lambda$1640 emitter
located at 3 pkpc from the galaxy GN-z11, at z=10.6. The detection, based on JWST NIRSpec-IFU high-resolution spectroscopy, confirms a previous claim based on medium-resolution spectroscopy. The HeII$\lambda$1640 identification is further supported by the independent detection of H$\gamma$ obtained by \"Ubler et al. (2026) at the same location. The HeII emission is spectrally resolved in two components separated by 120~km/s. The Equivalent Width of the HeII emission is extremely high ($>$20\AA). No metal lines are detected.
We argue that Population III stars are the most plausible explanation for the observed He II emission, with no satisfactory alternative from other classes of sources or mechanisms.
\end{abstract}

\begin{keywords}
galaxies: high-redshift -- galaxies: evolution -- galaxies: abundances -- stars: Population III -- infrared: galaxies
\end{keywords}



\section{Introduction}
\label{sec:intro}

The detection and characterization of the first population of stars, formed out of pristine clouds of gas, the so-called Population III (PopIII) stars, is one of the most important goals of modern astrophysics. These stars are theorised to form in galactic embryos within the first few hundred million years after the Big Bang \citep[e.g., ][]{Abel+2002,Yoshida+2003,Klessen:2023}.
There have been detections of galaxies with extremely low metallicities, in the range $\sim 10^{-3}-10^{-2}~Z_\odot$ \citep[e.g.,][]{Vanzella2023, Fujimoto:2025, Hsiao2025_lowZ, Maiolino:2025,Morishita2025_lowZ, Nakajima2025_lowZ}. These could be very low metallicity PopII galaxies, or possibly self-polluted or hybrid PopIII galaxies \citep[][]{Rusta:2025}. However,
these metallicities are still well above the expected threshold for PopIII, which are generally identified below a critical metallicity $Z_{cr} \sim 10^{-4}-10^{-6} Z_\odot$, which prevents the formation of low-mass stars because of inefficient gas cooling \citep[e.g.,][]{Bromm2002,Schneider2002,BrommYoshida2011_Review}. 

An alternative approach is to identify other tracers of PopIII. Most models expect PopIII stars to be very massive and very hot \citep[e.g.,][]{Trussler:2023, Schaerer:2025, Storck:2025, Wasserman:2026}, and their hard spectral energy distribution (SED) is expected to doubly ionize helium, resulting in significant emission of HeII recombination lines, such as $\rm He\textsc{ii}\lambda1640$ and $\rm He\textsc{ii}\lambda4686$ \citep[e.g.,][]{Schaerer2003, Zackrisson2011, Nakajima2022, Lecroq2025, Schaerer:2025, Storck:2025, Wasserman:2026}.

While most of the PopIII stars are expected to form in small haloes, various models have predicted that they may also form around massive haloes at high redshift as a consequence of gas accretion, inefficient mixing, and massive star formation resulting from gas compression and Lyman-Werner heating from the central source  \citep[e.g.,][]{Liu:2020, Venditti:2023, Venditti:2024, Venditti:2025}. Interestingly, \citet[][]{Venditti:2023} show that it is potentially more likely to find PopIII stars in distant massive haloes than in their low-mass counterparts.
While massive haloes are considerably rarer than their low mass counterparts in the early Universe \citep{Wechsler2018}, they typically host massive and luminous galaxies, which makes them easier to identify in observations.

Within this context, \citet[][]{Maiolino:2024_Halo} explored the immediate environment around the most UV-luminous galaxy at $z>10$, GN-z11 \citep[RA 12:36:25.44, DEC +62:14:31.3][]{Oesch:2016, Bunker2023, AlvarezMarquez:2025}. This $z=10.6$ galaxy has a stellar mass of $\sim 8\times 10^8~M_\odot$, is located in an overdense region \citep[][]{Tacchella2023, Scholtz:2024}, and likely hosts an AGN
\citep[e.g.,][]{Maiolino2024, Ji2025GNz11, Scholtz2025_type2, CrespoGomez:2026, Fabian:2026}. 
\citet[][]{Maiolino:2024_Halo} used NIRSpec-IFU medium resolution ($R \sim 1000$) spectroscopy to confirm a previous tentative detection of HeII$\lambda$1640 identified with the Prism ($R \sim 100$), along the MSA shutter targeting GN-z11. Specifically, they identified an emission line in a region located at about 3~kpc NE of GN-z11, at a wavelength consistent with HeII at a redshift very close to GN-z11 ($z\sim 10.600$). Given the inferred high equivalent width (EW) of the line ($>$20\AA) and the absence of metal lines, \citet[][]{Maiolino:2024_Halo} suggested this to be a star-forming clump with possible signatures of PopIII stars.

In this paper, we report deep high-resolution (G235H; $R \sim 2700$) NIRSpec IFU observations of the same region targeted in the previous study. These new observations confirm the presence of HeII emission in the halo of GN-z11. Further confirmation is obtained through the detection of H$\gamma$ at the same location and same redshift, which is reported in a companion paper \citep{Ubler2026}.

A second companion paper, \citet{Rusta2026}, presents a theoretical interpretation of our findings on this source in the Pop III scenario, also discussing the implications for the still-unknown mass distribution of Pop III stars.

Throughout this paper, and in the companion papers, this source is named
{\it Hebe}\footnote{HElium Balmer Emitter. In ancient Greek mythology Hebe (\accdasia{}\acctonos{}H$\beta \eta$) is the goddess of youth, daughter of Zeus and Hera.}.

A flat $\Lambda$CDM cosmology is adopted throughout based on the latest results of the Planck collaboration \citep{Planck2020}, with $H_0 = 67.4 \, \mathrm{km \, s^{-1} \, Mpc^{-1}}$, $\Omega_\text{m} = 0.315$, $\Omega_\text{b} = 0.0492$.

\section{Observations and data processing}
\label{sec:obs}

Observations were obtained through the JWST programme ID  (PI: Maiolino)  using the NIRSpec IFU 
\citep{Jakobsen2022,boeker2022,Boker2023}
with the high resolution grating G235H (R$\sim$2700), covering
a field of view of $3''\times 3''$. The observations were performed between May 15 and 17, 2025, split into two visits. Detectors were read out with the NRSIRS2 pattern, with 25 groups and 2 integrations.
A total of 39 dithers were executed, with a medium cycling pattern. Unfortunately, during dither position 18 the telescope guiding camera locked onto a hot pixel; hence, this frame had to be discarded in further processing. After removing this frame, the total exposure time was 38.8 hours.  
The data reduction uses the JWST Science calibration pipeline v1.15.0 with CRDS jwst\_1293.pmap. To improve the quality of the datacube, we performed a number of further data processing steps, in addition to the default pipeline. The customised procedures used for residual cosmic rays snowballs flagging, subtraction of pink noise (1/f noise), and removal for failed open MSA shutters signal are described in detail in \cite{Perna2023}. Stage~2 was executed using the bad-pixel self-calibration implemented in the standard pipeline, in which all exposures on a given detector are used to identify and flag bad pixels that may have been missed by the reference bad-pixel mask. 
For the final stage of the pipeline, outlier rejection was performed following \cite{DEugenio2024_pablo}, with a rejection threshold at the $98^{\rm th}$ percentile. Before combination, the cube was resampled to a scale of $0.05''$ per spaxel. 

The error extension of the cube is known to underestimate the actual error. Therefore, following \cite{Ubler2023}, we have rescaled the error extension to match the measured noise in regions of the cube free of sources.

The cube was registered with the NIRCam images by producing pseudo broad-band maps, by collapsing the cube across the large wavelength ranges matching three NIRCam filters. Unfortunately, only two sources in the field are available, GN-z11 and a relatively large foreground galaxy at $z=2.028$. This, in addition to the coarse pixel size of NIRSpec, results in an accuracy of the resulting registration of about 0.05$''$.

\begin{figure}
	\centering
\includegraphics[width=\linewidth]{"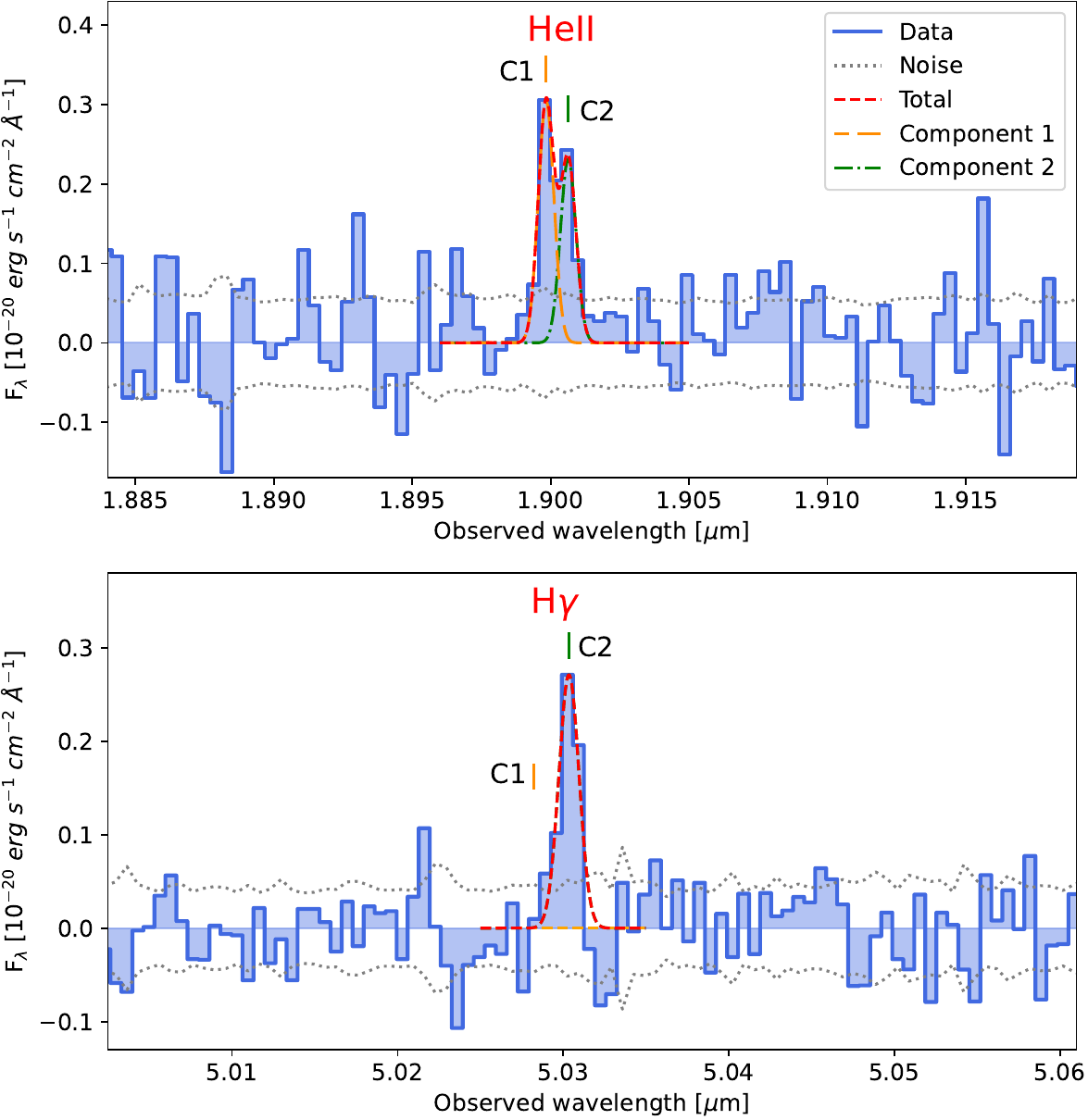"}
	\caption{
    Top: {\it Hebe}'s spectrum around the wavelength of HeII$\lambda$1640.  Bottom: {\it Hebe}'s spectrum around H$\gamma$, from \citet{Ubler2026}. The long-dashed (orange) and dot-dashed (green) lines show the simultaneous fit of HeII and H$\gamma$ with two components (C1 and C2), respectively, while the short-dashed red line shows the total; only C2 is detected in H$\gamma$. The flux scale is not corrected for aperture losses nor for lensing magnification.
	}
\label{fig:heii_spectrum}
\end{figure}

\section{{\it Hebe}: confirmation of a HeII emitter in the halo of GN-z11}
\label{sec:heii}

The upper panel of Fig.\ref{fig:heii_spectrum} shows the spectrum extracted from a $0.1''\times 0.1''$ aperture centred at $0.50''$ East and $0.55''$ North of GN-z11 (RA 12:36:25.520, DEC +62:14:31.934). The spectrum shows an emission feature, detected at 6$\sigma$, which is identified as HeII$\lambda$1640 at $z_{HeII}=10.583$ (centroid of the line), blue-shifted by about 450~km/s relative to GN-z11. Note that, for this paper, the redshift of GN-z11 is consistently remeasured based on its HeII emission in our cube, and found to be $z_{GN-z11,R2700}({\rm HeII-based})=10.6018$; this is slightly lower than the redshift given by \citet{Bunker2023} using medium resolution spectroscopy - $z_{GN-z11,R1000}=10.6034$- but it is also slightly higher than the redshift that would be derived from their HeII medium resolution detection, which would give $z_{GN-z11,R1000}({\rm HeII-based})=10.5975$; these slight inconsistencies are likely due to the lower spectral resolution of the previous observations.

This independently confirms the previous HeII detection at the same location obtained with medium resolution spectroscopy by \citet{Maiolino:2024_Halo}.

As already mentioned, we name {\it Hebe} this HeII emitting source, which is located at 3~kpc from GN-z11.

\citet{Maiolino:2024_Halo} already pointed out that no other line identification was plausible, because of the absence of other nebular lines, and that the HeII$\lambda$1640 line at a redshift very close to GN-z11 was the most likely explanation. Any possible ambiguity with potential lower redshift interlopers is now completely removed, as the HeII identification is unambiguously and independently confirmed by the detection of H$\gamma$, at the same location and redshift of {\it Hebe}, as reported in a companion paper by \citet{Ubler2026}.
The H$\gamma$ detection is shown for convenience in the bottom panel of Fig.\ref{fig:heii_spectrum}, while its thorough analysis is discussed in \citet{Ubler2026}. 

\begin{figure}
	\centering
\includegraphics[width=\linewidth]{"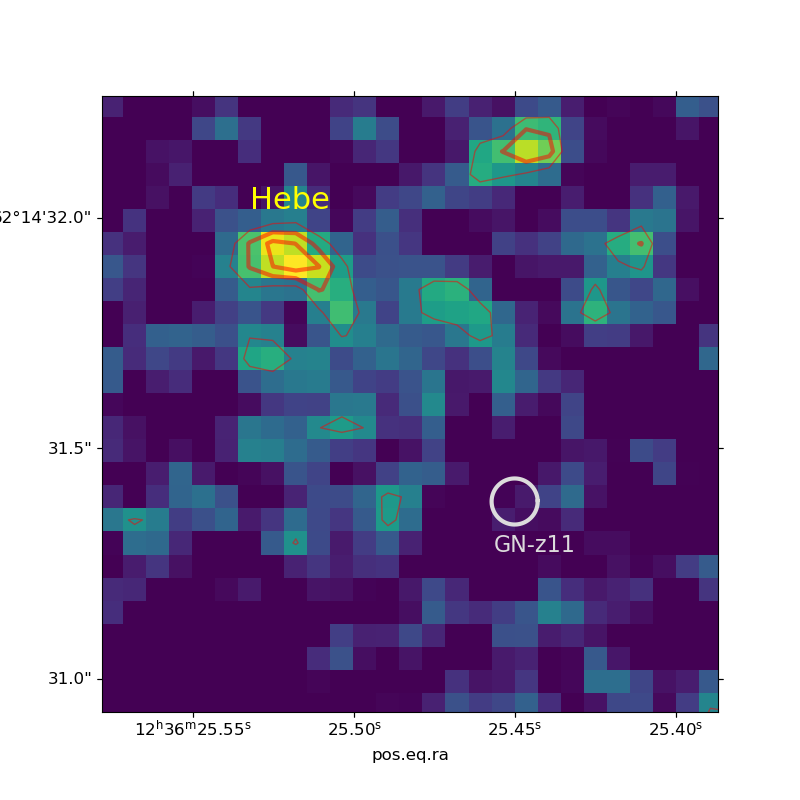"}
	\caption{Continuum-subtracted map of the HeII emission at the redshift of {\it Hebe} in the halo of GN-z11 (see text). Contours indicate the 3$\sigma$ (thin), 4$\sigma$ and 5$\sigma$ (thick) levels. The white circle indicates the location of GN-z11.
	}
\label{fig:heii_map}
\end{figure}

\begin{figure}
	\centering
\includegraphics[width=0.8\linewidth,clip, trim=6.5cm 0.0cm 8.5cm 0.0cm]{"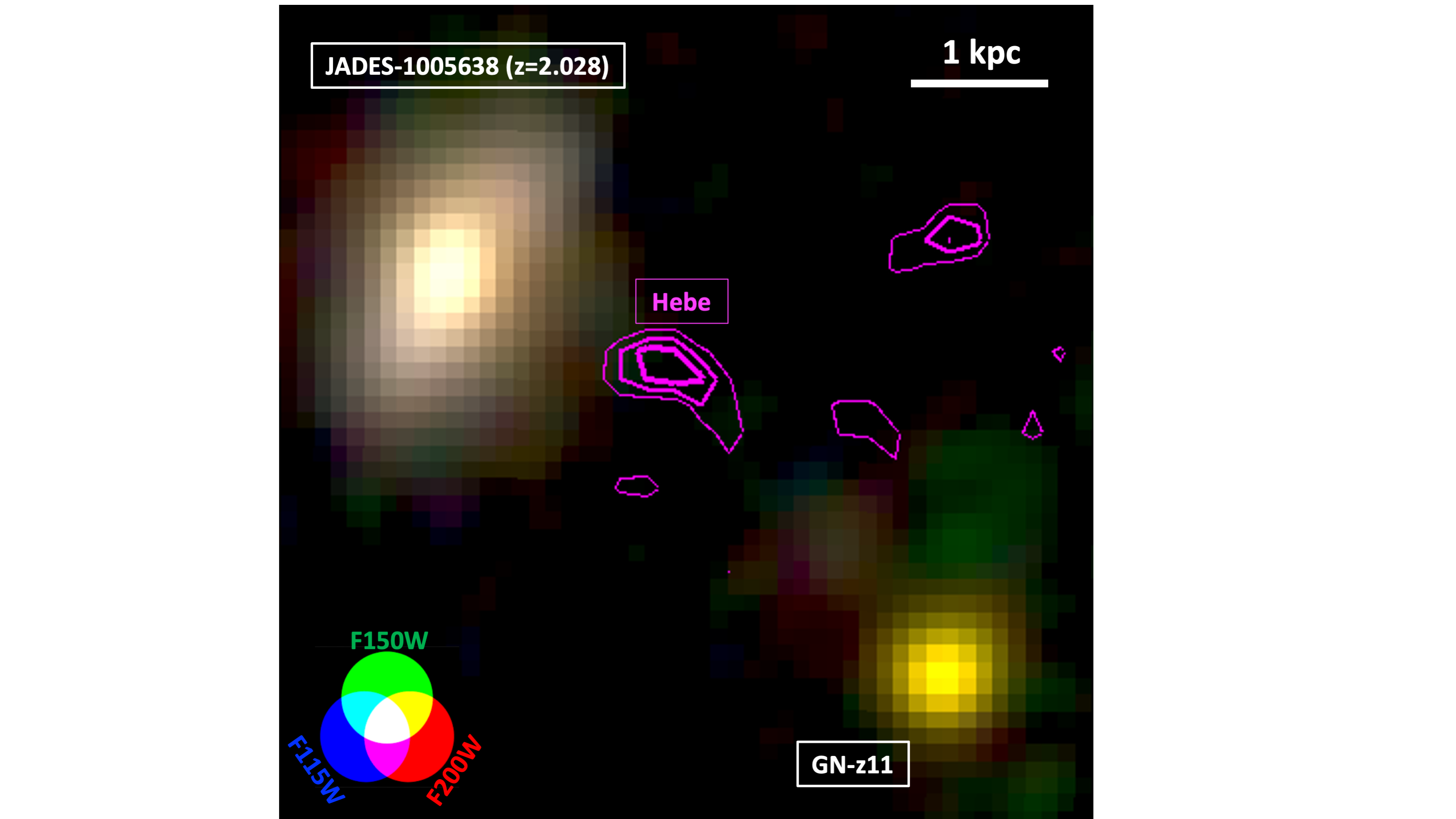"}
	\caption{Same map of the HeII emission as in Fig.\ref{fig:heii_map} (contours), overlayed on the RGB image of the field from the NIRCam filters F115W, F150W and F200W, and with a $1.5\arcsec\times1.5\arcsec$ Field of View, adjusted to show the location of both GN-z11 and the foreground galaxy at z=2.03.
	}
\label{fig:heii_map_RGB}
\end{figure}

Fig.\ref{fig:heii_map} shows the HeII map by collapsing the continuum-subtracted spectral channels between 1.8995~$\mu$m and 1.9007~$\mu$m (the three central channels of the line, i.e. a velocity interval of 187 km/s), which highlights the location of  {\it Hebe}. The white circle indicates the location of GN-z11. Fig.\ref{fig:heii_map_RGB} shows the HeII contours again, but overlaid on the NIRCam RGB image, with the Field of View adjusted to show the location of {\it Hebe} relative to GN-z11 and a foreground galaxy at z=2.04, which causes some gravitational lensing (see Appendix \ref{lens_app}).

As discussed in more detail in Section \ref{sec:mediumR} below, the location and flux of the HeII emission are broadly consistent with the previous detection based on the NIRSpec medium resolution grating \citep{Maiolino:2024_Halo}, although the comparison with the latter suggests the presence of some diffuse emission, both spatially and in velocity. Fig.\ref{fig:heii_map} shows the compactness of the line emission, which is unresolved  ($<$ 400 pc in size) at the NIRSpec resolution. We note some diffuse emission extending towards the SW. Even though its surface brightness is not statistically significant (mostly within $<2\sigma$ of the background fluctuations), a similar diffuse emission was seen in the medium resolution map. This will be discussed again in Section \ref{sec:mediumR}. There is also another marginally detected emitter 0.8$''$ North of GN-z11; this was also marginally detected in medium resolution map, and will be discussed in Section \ref{sec:mediumR}.

The HeII emission has a  FWHM of
about 110 km/s (deconvolved by the spectral resolution)\footnote{Following, e.g., \citealp{DEugenio2025} or \citealp{Shajib:2025}, we take the spectral resolution $R$ as the nominal value \citep{Jakobsen2022}, but at 1.9~$\mu$m this is divided by a factor of 0.7 (determined empirically in those references), which takes into account that, for compact sources, the size of the source can be comparable to the size of the IFU slices.}, but this apparent width is actually resulting from two marginally resolved components. This is in contrast with the H$\gamma$ emission, which is spectrally unresolved (see \citealt{Ubler2026}; note that the two spectra have similar resolution -- $R\sim 2860$ at 1.9$\mu$m and $R\sim 3500$ at 5$\mu$m). We have simultaneously fitted HeII and H$\gamma$ with two Gaussian components (C1 and C2), as shown by the dashed lines in Fig.\ref{fig:heii_spectrum}. The two components are separated by $126\pm 17$~km/s, and each of them is consistent with being spectrally unresolved ($\sigma_V <35~km~s^{-1})$.
The simultaneous fit indicates that H$\gamma$ is associated only with C2, while the component C1 of H$\gamma$ is undetected.

We have also attempted a simultaneous fit with a single Gaussian component, which results in a significantly worse fit, as shown in Appendix \ref{app:oneGfit}.

It is difficult to explore whether there is any small spatial offset between the two components C1 and C2, as the maps of individual spectral channels become too noisy for a proper assessment.

The inferred fluxes and luminosities of the HeII line and its two individual components are reported in Table~\ref{tab:fluxes}. These include an aperture correction factor of 3.03, estimated from the cubes of two stars (ID 1219 --PI: Luetzgendorf -- and ID 1222 --PI: Willott--) processed with the same methodology, at the wavelength of 1.9$\mu$m. 
The total flux has been estimated by simply summing up the flux in the central 5 spectral pixels of the line. The flux of the individual components is determined by the simultaneous double Gaussian fit.

We also correct the luminosities for a magnification of 1.42 due to the foreground galaxy at z=2.028. Details of the lensing model and magnification estimation are given in Appendix \ref{lens_app}.

As discussed in Section \ref{sec:mediumR}, there is evidence for a diffuse component of HeII; in the second part of Table \ref{tab:fluxes} we also provide an estimate of {\it Hebe}'s HeII flux after subtracting this more diffuse component\footnote{This is done by taking the diffuse flux estimated in Sect.\ref{sec:mediumR} and deriving the contribution to the extraction aperture.}.

\begin{table}
    \centering
    \begin{tabular}{lc}
    \hline
    \multicolumn{2} {l} {
    Aperture corrected fluxes} \\
    \hline
        \hline
    \multicolumn{2} {l} {
    Total} \\
    $F(HeII1640)$ & $(1.11 \pm  0.17)\times 10^{-19}$ $\rm erg~s^{-1}~cm^{-2}$ \\
    $EW_0(HeII1640)$ & $>47$\AA \\
    $NIV]_{1483,1486}/HeII$ & $<  0.60^a$ \\
    $CIV_{1548,1551}/HeII$ & $<  0.56^a$ \\
    $OIII]_{1661,1666}/HeII$ & $<  0.63^a$ \\
    $NIII]_{1749}/HeII$ & $<  0.46^a$ \\
    $CIII]_{906,1908}/HeII$ & $<  0.36^a$ \\
    $F200W$ & <3.1 (<2.7)$^b$ nJy \\
    \hline
        \multicolumn{2} {l} {
    Component 1} \\
    $F(HeII1640)$ & $(6.7 \pm  1.2)\times 10^{-20}$ $\rm erg~s^{-1}~cm^{-2}$ \\
    $EW_0(HeII1640)$ & $>28$\AA \\
    \hline
    \multicolumn{2} {l} {
    Component 2} \\
    $F(HeII1640)$ & $(5.1 \pm  1.1)\times 10^{-20}$ $\rm erg~s^{-1}~cm^{-2}$ \\
    $EW_0(HeII1640)$ & $>22$\AA \\
        \hline
        \hline
        \multicolumn{2} {l} {
    De-lensed luminosities and cleaned of diffuse component } \\
    \hline
    $L(HeII1640)$ (total) & $(8.54 \pm  1.29)\times 10^{40}$ $\rm erg~s^{-1}$\\
    $L(HeII1640)$ (C1) & $(5.1 \pm  0.9)\times 10^{40}$ $\rm erg~s^{-1}$\\
    $L(HeII1640)$ (C2) & $(3.9 \pm  0.9)\times 10^{40}$ $\rm erg~s^{-1}$\\
    \hline
    \end{tabular}
    \caption{
    Fluxes and equivalent widths for {\it Hebe} and upper limits of the associated emission lines. 
    The fluxes in the upper section are corrected for aperture losses. The luminosity in the bottom section are also corrected for a lensing magnification factor of 1.42 and also for the contribution of the diffuse component, as discussed in the text. Upper limits of all emission lines are at $3\sigma$. Notes: $^a$for all doublets/multiplets the table reports the upper limit on each individual component of the doublet, as they would be spectrally resolved; $^b$the F200W upper limit is at $5\sigma$, aperture-corrected for the NIRCam PSF -- the value in parentheses is corrected for the HeII line contribution to F200W.
    }
\label{tab:fluxes}
\end{table}

\section{Equivalent width constraints}
\label{sec:ew}

Measuring or constraining the equivalent width of the HeII emission is incredibly important, as it can provide key constraints on the nature of the source. This requires a detection of the continuum emission. For extremely low metallicity galaxies, the main difficulty is the continuum detection rather than the nebular line emission. A clear example is the extremely metal poor galaxy LAP-1 at $z=6.6$ \citep[][]{Vanzella2023}, where multiple nebular emission lines are clearly detected, while the continuum remains undetected despite deep exposures and lensing magnification  \citep[][]{Nakajima2025_lowZ}. The HeII emitter in GN-z11 has a similar issue, with the additional complication that it is in the proximity of the relatively extended foreground galaxy at z=2.028. 
As discussed in \citet[][]{Maiolino:2024_Halo} \citep[see also][]{Tacchella2023}, modelling and removing the emission from the foreground galaxy leaves some residual flux around the location of {\it Hebe}. Yet, there is no clear indication of a continuum point source associated with it, suggesting that the flux is some diffuse residual after the foreground galaxy removal. Thanks to the more precise localization of {\it Hebe}, we can now confirm the absence of a continuum detection at the 5$\sigma$ upper limit level of 3.1~nJy in F200W. We derive this value from the same aperture adopted for the spectrum extraction, accounting for a correction for a point-like source. Direct photometry (also corrected for the limited aperture size) provides a flux measurement of $0.8\pm0.6$~nJy.
The 5$\sigma$ upper limit was corrected for the contribution of the HeII line to the broad-band measurement to get a spectral continuum upper limit. The rest-frame EW(HeII) corresponding to this continuum is 47~\AA.
We provide a conservative lower limit on the EW of the C1 and C2 components by applying the same continuum upper limit to each of them. This translates into lower limits of 28~\AA \ and 22~\AA, for the C1 ad C2 components, respectively. All these measurements are provided in Table~\ref{tab:fluxes}.

As we discuss extensively in Section~\ref{sec:PopIII}, the inferred rest-frame EW(HeII) are higher than 20\AA, well above the EW expected for a metal-enriched (PopII) stellar population. If the EW(HeII) were $<20$\AA \ then we would have expected the $\sim 10 ~\sigma$ detection of a source with a flux higher than 6~nJy in F200W.

\section{Absence of metal emission lines}
\label{sec:nometals}

The G235H/F170LP setup covers a wavelength range where various metal emission lines typically detected in high-redshift galaxies \citep[e.g.][]{DEugenio2024,Scholtz2025DR4} are expected to be seen at the redshift of {\it Hebe}. Specifically, the NIV]1483,1486, CIV1548,1551,  OIII]1661,1666, and CIII]1906,1908 doublets, as well as the NIII]1749 multiplet, are all within the spectroscopic range of this setup.
Considering that in local and high-z galaxies at least some these lines (if not all of them) are stronger than HeII, at least some of these metal lines should be visible also in {\it Hebe}.
On the contrary, none of these transitions is detected in {\it Hebe}. We report conservative $3\sigma$ upper limits of the flux of these metal lines relative to HeII. The spectral regions around the location of these lines are shown in Appendix \ref{app:oneGfit}.

Moreover, in the companion paper by \citet{Ubler2026} we show that no lines are detected in the G395H spectrum aside from the prominent H$\gamma$ (and a tentative detection of H$\delta$). In particular, we do not detect [NeIII]3869, whose strength is often comparable to H$\gamma$ in high-z galaxies \citep{Scholtz2025DR4}.

The companion paper by \citet{Rusta2026} provides a theoretical interpretation of our observational findings, showing that these non-detections are very constraining of the nature of the ionizing sources.

\begin{figure}
	\centering
\includegraphics[width=\linewidth]{"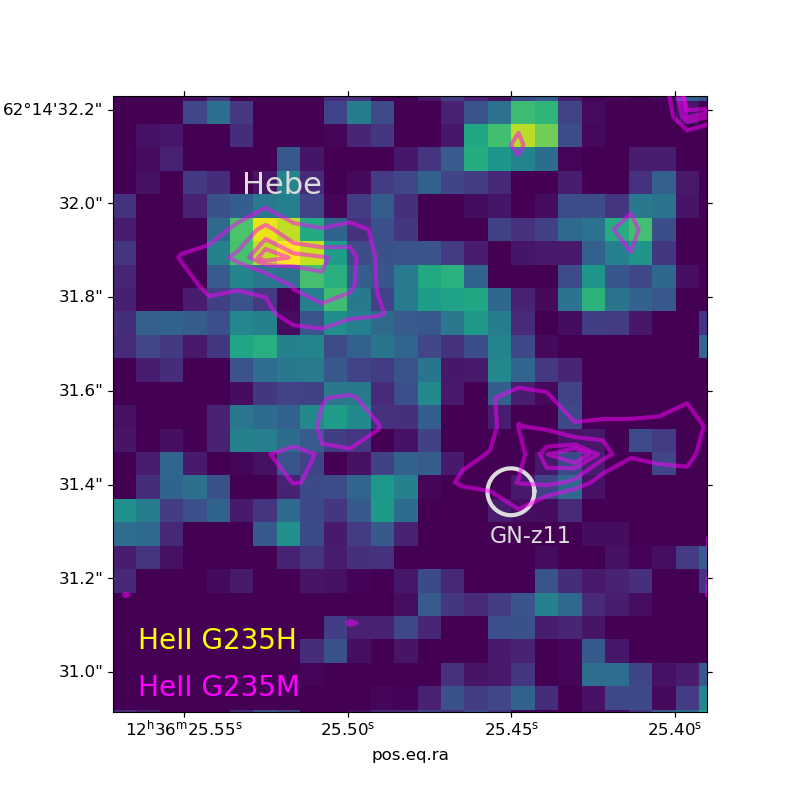"}
	\caption{Overlay of the continuum-subtracted medium resolution R1000 map of HeII reported in \citet[][]{Maiolino:2024_Halo} (magenta contours) on the continuum-subtracted R2700 map of HeII obtained in this paper (background image). The white circle indicates location of the continuum of GN-z11 and its size shows the relative uncertainty of positioning the two continua for registering the two maps. Note that at R1000 emission is also seen close to GN-z11, as at low resolution the broader integrating wavelength window includes some HeII emission from GN-z11. 
    Also note that in the R1000 map the top-left corner is affected by an artefact originating from the fact that, in that observation, Westward of GN-z11, there were fewer frames and edge effects, due to a guiding problem \citep[see][for details]{Maiolino:2024_Halo}.
    }
\label{fig:heii_R1000_R2700}
\end{figure}

\section{Consistency with the medium resolution observation and diffuse HeII emission}
\label{sec:mediumR}

In this section, we compare the HeII detection in our new high-resolution (G235H) data to the previous detection in the medium resolution spectrum (G235M) presented in \citet[][]{Maiolino:2024_Halo}.

The aperture-corrected HeII flux of {\it Hebe} (and not corrected for magnification) is $(1.11\pm 0.18) \times 10^{-19}~\rm erg~s^{-1}~cm^{-2}$. This is to be compared with the flux of $(1.8\pm 0.34) \times 10^{-19}~\rm erg~s^{-1}~cm^{-2}$ reported by \citet[][]{Maiolino:2024_Halo}. The two fluxes, therefore, are marginally consistent, within $\sim 2\sigma$. However, the comparison is more complex, as discussed in the following.

To begin with, we compare the HeII spectra of GN-z11 in the two cubes and notice that there is probably a slight wavelength miscalibration between the two, in the sense that the G235M spectrum requires a blueshift by about 0.6 spectral pixels to be aligned with the G235H spectrum. It is not clear whether this is due to a  potential issue in the calibration, or to different versions of the calibration files. However, investigating these aspects is beyond the scope of this paper, and we simply apply the slight wavelength shift to the G235M spectrum. 

Secondly, in Fig.\ref{fig:heii_R1000_R2700} we compare the R2700 HeII map obtained in this work, with the R1000 HeII map obtained in \citet[][]{Maiolino:2024_Halo} (contours). Note that the G235M map shows significant HeII emission associated with GN-z11, not seen in the G235H map because the velocity range integrated in G235M is larger and includes also some HeII emission from GN-z11 (recall that {\it Hebe} is blueshifted by $\sim 450$~km/s relative to GN-z11).
We have astrometrically registered the two images via the GN-z11 continuum. Unfortunately, this leaves some uncertainty in the relative positions. Specifically, in addition to the coarseness of the NIRSpec spatial sampling ($0.1''$ native spaxels), 
 there are only two continuum sources that can be used for the alignment: GN-z11 and the foreground galaxy, which is not point-like but fairly extended.
The white circle in the figure illustrates the location of GN-z11, and its size illustrates the uncertainty in aligning the two continua. The HeII emission in G235H appears slightly offset towards the NE relative to the emission seen in G235M, although within the maps relative alignment uncertainty. It is also true that the G235M emission is not symmetrically distributed and its peak corresponds to the maximum emission in G235H, within half a (resampled) pixel. It is possible that there is a residual alignment issue, but it is also possible that the two maps do not match exactly because they are probing slightly different velocity ranges, as discussed below.

The HeII flux reported in \citet[][]{Maiolino:2024_Halo}, was integrated on a larger velocity range (simply because of the lower resolution and larger spectral channels), and also in a larger aperture ($0.24''\times 0.24''$), centred more towards the South relative to the smaller ($0.1''\times 0.1''$) aperture used in this paper. We therefore re-extract a spectrum on an aperture and velocity range matching the extraction of \citet[][]{Maiolino:2024_Halo} -- the spectrum is much noisier, but the total flux is $(1.50\pm 0.45)\times 10^{-19}~\mathrm{erg~s^{-1}~cm^{-2}}$, hence matching very well the flux observed in G235M. This indicates that there is some extended HeII flux, diffuse both spatially and in velocity, that is probably not clearly visible in the high-resolution data due to the higher noise affecting the signal distributed on larger velocities. Some of this diffuse emission is actually potentially detected in the HeII maps of Figs.\ref{fig:heii_map} and \ref{fig:heii_R1000_R2700}, and was potentially seen also in the medium resolution map of \citet[][]{Maiolino:2024_Halo}. The origin of such diffuse emission, probably with high velocity (of the order of a few hundreds km/s), is not clear. However, we note that \citet[][]{HamelBravo:2025} detected broad (655~km/s) HeII emission (without corresponding broad Balmer emission) in the halo of the ultra-metal poor local galaxy SBS~0335-052E; the nature of such broad HeII emission is uncertain, but could result from near-pristine gas shocked to high temperature by the outflow driven by the central galaxy. A similar scenario could be happening in the vicinity of {\it Hebe}: the vigorous star-forming activity may have driven an outflow that is shock-heating the gas to high temperatures. However, the inferred velocities are quite modest ($\sim 200$~km/s), so it is unlikely that shocks are relevant \citep{Lecroq2024}. Additionally, we warn that the presence of high-velocity gas is inferred only indirectly from the total flux in the large aperture, and when compared with the medium-resolution data, deeper data would be needed to directly confirm it.

We finally note that some diffuse emission was identified in the NIRCam images in the vicinity of GN-z11, within a radius of about $0.25''$ North of GN-z11, the so-called `Haze' \citep{Tacchella2023}. This is marginally visible in Fig.\ref{fig:heii_map_RGB}. The origin of this emission is not clear, but \citet{Tacchella2023} suggests that it is a foreground, low-redshift interloper. Here we simply notice that such `Haze' does not seem to be physically associated with the tentative diffuse emission seen in HeII.

\begin{figure*}
	\centering
\includegraphics[width=\linewidth]{"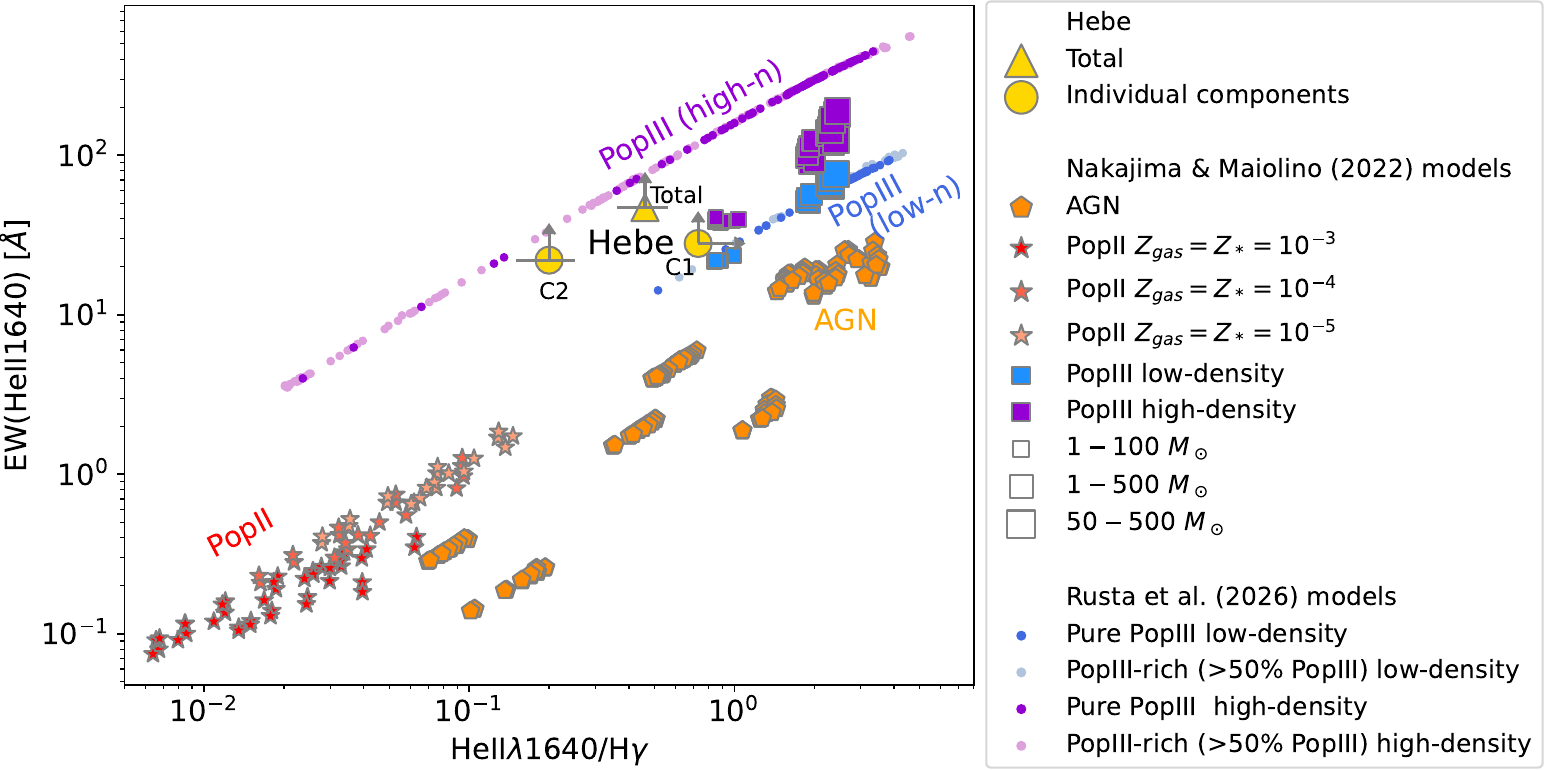"}
	\caption{EW(HeII1640) versus HeII/H$\gamma$ diagnostic diagram. Various symbols show models from \citet[][]{Nakajima2022} for different classes of objects, specifically: squares - PopIII (blue: density $n=10^3~\rm cm^{-3}$; purple: densities $n=10^5~\rm cm^{-3}$ and $10^6~\rm cm^{-3}$); red stars - PopII with decreasing metallicity, from darker to lighter, as indicated in the legend; orange pentagons - AGN. The sizes of the PopIII symbols reflect different IMFs. The small points are pristine PopIII models from \citet{Rusta2026}, from pure PopIII (dark symbols) to cases with PopII contribution but still PopIII-dominated (PopIII $>$50\% in mass, light symbols). The large golden symbols are the values inferred for {\it Hebe}, both the total and individual components, as indicated.
	}
\label{fig:diag}
\end{figure*}

\section{The Population III scenario}
\label{sec:PopIII}

Spectra with prominent HeII emission and absence of metal lines are expected in the case of Population III stars \citep[e.g.][]{Zackrisson2011,Trussler:2023}. In this section, we explore this scenario more quantitatively through a simple comparison with some models and expectations.

Fig.\ref{fig:diag} shows the EW(HeII$\lambda$1640) vs HeII$\lambda$/H$\gamma$ diagnostic diagram. This diagram was already presented in \citet[][]{Maiolino:2024_Halo} for {\it Hebe}. In that paper, however, H$\gamma$ was measured along the shutter using the prism, hence likely only grazing at the more diffuse emission and not probing the core of  {\it Hebe} (see discussion in Section \ref{sec:mediumR}). With the new data presented in \citet{Ubler2026}, we have a direct measurement of H$\gamma$ in the same region and with a similar spectral resolution as for the HeII emission. 
The various symbols show models from \citet[][]{Nakajima2022}, with various ionization parameters and metallicities. Given the complete absence of metal lines, to avoid overcrowding the plot, we focus on low metallicity models, below 10\% solar. Specifically, 
red stars indicate PopII models, with decreasing metallicity from dark to light, down to a metallicity of $10^{-5}$ (i.e. $7\times 10^{-4}~ Z_\odot$). Note that these models use BPASS binary stellar templates, which are already more energetic than standard stellar populations
\citep[see][for details]{Nakajima2022}.
Orange pentagons are AGN models, with a broad range of possible ionizing spectra, a broad range of ionizing parameters, and a very broad range of metallicities, from 10\% solar to completely pristine.  
The squares show the PopIII models, with sizes indicating the adopted IMF (assuming a Salpeter slope): larger symbols are for stellar masses in the range $50-500~M_\odot$, intermediate size symbols are for stellar masses in the range $1-500~M_\odot$, and small symbols for stellar masses in the range $1-100~M_\odot$. The blue squares are for gas densities of $10^3~\rm cm^{-3}$. The purple squares are for gas densities of $10^5~\rm cm^{-3}$ and $10^6~\rm cm^{-3}$.
The small blue points are PopIII models from  \citet[][]{Rusta2026}: dark points are for Pure III pristine systems, while light points are for systems with some PopII contamination, but still dominated by PopIII ($>$50\%). The blue points are for gas densities of $10^3~\rm cm^{-3}$ and ionization parameter $U=0$, the purple points are for gas densities of $10^6~\rm cm^{-3}$ and ionization parameter $U=-2$.

We note that densities of $n\sim 10^5-10^6~\rm cm^{-3}$ explored in the models may appear somewhat extreme; however, high redshift studies have revealed a significant population of galaxies with gas densities in this range \citep[e.g.][]{Martinez2025,Moreschini2026}. Therefore, it is very plausible that PopIII systems can have such high densities.
The reason why PopIII models with high densities are significantly higher in EW is that these are above the critical density of the 2-photons nebular continuum, which is therefore suppressed, boosting the EW(HeII). Interestingly, the same does not happen for AGN; indeed, in these cases, the 2-photon continuum is negligible, and the UV continuum is dominated by the accretion disc. Therefore, increasing the gas density to $n\sim 10^5-10^6~\rm cm^{-3}$ for the AGN models does not have any tangible effect.

The large golden symbols are the values measured for {\it Hebe}, both total (triangle) and for the individual components (circles), as labelled. 
The EW(HeII) and HeII/H$\gamma$ ratios observed for {\it Hebe}, either total or for the individual components, are well above the expectations for PopII models. AGN models also struggle to reproduce the observed constraints, primarily because they generally predict lower EW(HeII); however, we will discuss in more detail the AGN scenario in section \ref{sec:dcbh}. Component C1 is the one most clearly aligned with the PopIII prediction, both for the low- and high-density models. Component C2 seems inconsistent with 
PopIII low-density models, but still consistent with PopIII high-density models.
We shall mention that it is possible that other PopIII models, with steeper SED and higher yield of HeII photons, which for example are associated to rotating or binary PopIII stars \citep[e.g.][]{Yoon2012,Murphy2021, Sibony2022,Wasserman2026,Lecroq2025}, could reproduce C2 with low density gas. These models were not considered in
\citet{Nakajima2022} nor in \citet{Rusta2026}. An alternative possibility is that in C2 the very first generation of PopIII may have provided a first quick enrichment of dust, as expected by many models \citep[see review by][]{Schneider:2024}. Taking the extinction curve by \citet{Sun2026} (one of the few extinction curves measured at high redshift), an extinction of only $A_V\sim 0.25$ would be enough to move the HeII/H$\gamma$ flux ratio to the PopIII sequence. Depending on the dust distribution geometry, in a compact configuration such an extinction can be achieved with just a few solar masses of dust \citep[see e.g.][]{Madau_Maiolino_2026}.

Overall, both for C1 and C2 excitation by PopIII stars appears to be the scenario most consistent with the current observational constraints on {\it Hebe}. In this scenario, the two components are two nearby clusters (within 400 pc of each other), probably at slightly different evolutionary stages (possibly with C2 being already contributed by some PopII) and/or slightly different local physical properties (different densities and ionization parameters).

A more extensive comparison with more detailed models is provided in the companion paper by \citet{Rusta2026}, who reach similar and even more stringent conclusions. They also constrain the total  stellar mass of the PopIII star clusters in the range between $10^4~M_\odot$ and a few times $10^5~M_\odot$. We note that $10^4~M_\odot$ is already well in the range expected for PopIII systems \citep[e.g.][]{Storck:2025}, although possibly towards the upper end, which is not unexpected given the peculiar environment. One should also take into account that there are various uncertain factors that could result in further reducing the inferred stellar mass, specifically: i) the lensing factor of the foreground galaxy is uncertain, and could result in a mass two times lower; ii) as mentioned, other PopIII models (especially rotating and/or binary PopIII stars) can result higher luminosity for a given mass, which would alos result in a reduced mass; iii) as we have spectrally resolved {\it Hebe} in two components, it is likely that also these two components are made of even lower mass systems, whose individual spectral signatures are blended within our spectral resolution.

We finally discuss that, as already mentioned in Section \ref{sec:heii}, there is another secondary peak in the G235H HeII map, located at about 0.8$''$ North of GN-z11. Based on the R2700 alone, this would certainly not be significant. However, it is interesting that it is marginally detected at the same location also in the G235M map, which may support that this is a second HeII emitter. If so, this would fit a scenario in which the halo of GN-z11 has recently been subject to pristine gas accretion from the N-NE, which is resulting in PopIII star formation at various locations as a consequence of gas compression. We note that there is possibly a third peak to the NW of GN-z11 seen both in the G235H and G235M maps; however, this is weaker in G235H and is also located in the region of G235M (West of GN-z11) where there are fewer frames and possible edge effects, hence less reliable.

\begin{figure}
	\centering
    \includegraphics[width=\linewidth]{"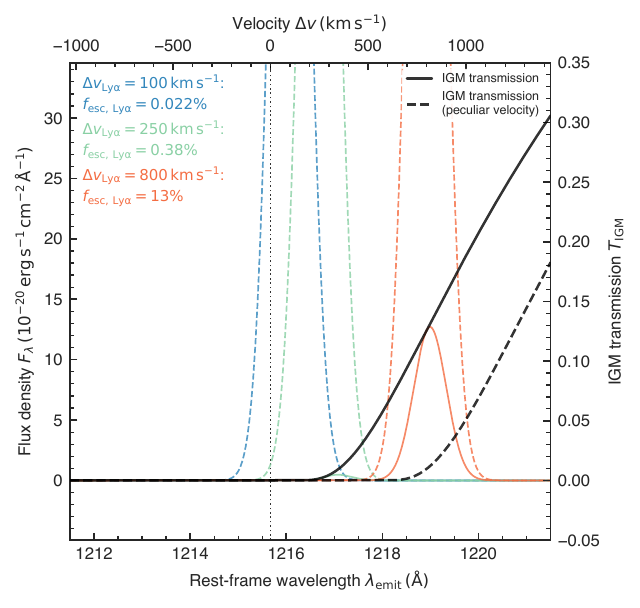"}
	\caption{IGM transmission curve of a fully neutral IGM at $z = 10.586$ (solid black line). Dashed colored lines show model Ly$\alpha$ line profiles of {\it Hebe} assuming different velocity offsets from systemic, normalized to the case-B expectation based on the H$\gamma$ line strength. Transmitted profiles are shown by solid lines, although only the case of $\Delta v_\text{Ly$\alpha$} = 800 \, \mathrm{km/s}$ has sufficient transmission for it to be clearly visible. Also shown is the case where {\it Hebe} has a peculiar velocity of $-450 \, \mathrm{km/s}$ relative to GN-z11 (dashed black line), which is expected to cause near-complete absorption of any Ly$\alpha$ emission.
	}
\label{fig:IGM}
\end{figure}

\section{\texorpdfstring{Lyman-$\alpha$}{Ly-alpha}}
\label{sec:Lya}

Most models of PopIII stars expect these to power strong Ly$\alpha$ emission \citep{2010A&A...523A..64R}. However, at $z=10.6$ the IGM absorption of Ly$\alpha$ is extremely strong, so that the vast majority of $z>9$ galaxies do not show any Ly$\alpha$ emission, despite often displaying prominent nebular emission lines \citep[e.g.][]{2024ApJ...975..208T, 2024A&A...683A.238J, 2025ApJS..278...33K, 2024MNRAS.535.1796B}. There are a few rare cases of Ly$\alpha$ emitters at high redshift \citep{2012ApJ...744...83O, 2013Natur.502..524F, 2015ApJ...804L..30O, 2015ApJ...810L..12Z, 2016ApJ...823..143R, 2022ApJ...930..104L, 2023MNRAS.526.1657T, 2024ApJ...970...50C, 2025MNRAS.536...27W, 2025arXiv251206072L}, the most remarkable one being JADES-GS-z13-1-LA at $z=13$ \citep{2025Natur.639..897W}. This galaxy has likely carved a small ionized `bubble' within the nearby IGM that allows its Ly$\alpha$ to escape. Despite its very high redshift, GN-z11 also shows the presence of some weak and extended Ly$\alpha$ emission  \citep{Bunker2023, 2024A&A...687A.283S}, although the presence of an ionised bubble is not strictly required. The Ly$\alpha$ emission in GN-z11 is significantly redshifted (by $\Delta v_\text{Ly$\alpha$} = 555 \pm 32 \, \mathrm{km/s}$), and much weaker than expected from the Balmer lines \citep[$f_\text{esc, Ly$\alpha$} = 0.038 \pm 0.004$;][]{Bunker2023}, indicating that it is in fact heavily suppressed by IGM damping-wing absorption.

An IFU G140M observation of GN-z11 covering the location of {\it Hebe} is available \citep{2024A&A...687A.283S}. This observation is much shorter than those with G235M or G235H. Ly$\alpha$ is not detected at the location of {\it Hebe}, with a $3\sigma$ upper limit of $15\times 10^{-19} \, \mathrm{erg \, s^{-1} \, cm^{-2}}$. This includes the aperture correction and is integrated over a velocity range of 150 km/s. Also, H$\gamma$ is 5 times weaker in {\it Hebe} compared to GN-z11, suggesting that the {\it intrinsic} Ly$\alpha$ luminosity of this regions is lower by a comparable factor \citep[PopIII and AGN are expected to have a similar HeII/Ly$\alpha$ intrinsic ratio, ][]{Nakajima2022}.

We have simulated IGM transmission curves at $z = 10.586$, the systemic redshift measured by \citet{Ubler2026}, using the \citet{2020MNRAS.499.1395M} model adopted in the \texttt{lymana\_absorption} code \citep{2025MNRAS.536...27W, 2025Natur.639..897W}\footnote{Publicly available at \url{https://github.com/joriswitstok/lymana_absorption}.}. We conservatively assume an intrinsic line FWHM of $200 \, \mathrm{km \, s^{-1}}$, almost twice that of H$\gamma$. We consider the IGM to be fully neutral at this redshift, since for GN-z11 itself this correctly reproduces the inferred Ly$\alpha$ transmission ($4\%$) at the observed velocity offset of the line \citep{Bunker2023}. As shown in Figure~\ref{fig:IGM}, for velocity offsets below $\Delta v_\text{Ly$\alpha$} \lesssim 250 \, \mathrm{km/s}$, appropriate for a faint, low-mass system \citep[$M_\text{UV} \gtrsim -20 \, \mathrm{mag}$; e.g.][]{2018ApJ...856....2M}, the predicted overall Ly$\alpha$ transmission is well below $1\%$, bringing the expected observed flux in agreement with the upper limit. Specifically, the expected Ly$\alpha$ flux would be $F_\text{Ly$\alpha$, obs.} < 10^{-19} \, \mathrm{erg \, s^{-1} \, cm^{-2}}$, i.e. more than an order of magnitude below the IFU upper limit.

The observed $3\sigma$ upper limit on Ly$\alpha$ would only be violated if the velocity offset were extremely high, $\Delta v_\text{Ly$\alpha$} \gtrsim +800 \, \mathrm{km/s}$, relative to the system rest frame (i.e. the Ly$\alpha$ tail extending even further redwards than what is seen in GN-z11). Additionally, we note that if {\it Hebe} is truly a satellite located within $\lesssim 10 \, \mathrm{pkpc}$ of GN-z11 as suggested by their close on-sky separation ($3 \, \mathrm{pkpc}$), the $-450 \, \mathrm{km/s}$ shift of {\it Hebe} relative to GN-z11 represents a peculiar velocity (rather than a cosmological redshifting effect, which would correspond to a line-of-sight separation of $0.30 \, \mathrm{pMpc}$). Compared to GN-z11, whose Ly$\alpha$ emission is already highly attenuated, the implied (additional) relative motion towards the neutral IGM along the line of sight would result in an even further suppression of the Ly$\alpha$ emission from {\it Hebe} (dashed black line in Figure~\ref{fig:IGM}).

Therefore, Ly$\alpha$ is essentially impossible to be detectable from {\it Hebe}, and this holds even for future, ultra-deep observations (of the order of 100 hours). The negligible  Ly$\alpha$ flux expected to reach us is thus fully consistent with our upper limit.

\section{The Wolf-Rayet Scenario}
\label{sec:wr}

One alternative interpretation of the observed HeII emission is the presence of a population of Wolf-Rayet (WR) stars. Specifically, in this scenario the broader HeII line, if not interpreted as two separate components, could be tracing a population of very metal-poor WR stars. 
Although broader than H$\gamma$, the  HeII line is much narrower (only 110~km/s) than typically observed in WR stars. Yet, 
the width of the HeII lines in the WR expanding envelopes is expected to decrease as a function of metallicity, as most of the radiation pressure generating the envelope winds happens via metal
line-driven winds. It is indeed observed that WR stars in more metal-poor environments tend to have narrower HeII
\citep{Vink2005,Crowther2006,Crowther2023,Castro2018}. However, even these low metallicity cases present prominent nitrogen (or carbon) emission lines, as (by definition of WR) the expulsion of the outer layers exposes the inner He and N burning core \citep[e.g.][]{Hainich2014,Berg2024,Berg2025}.

The WR phenomenon in the extremely metal-poor regime ($< 10^{-2}~Z_\odot$),
has been studied only theoretically.
The photospheric wind velocities are expected to decrease steadily as a function of metallicity \citep{Gafener2008,Sabhahit2023,Boco2025}, but even in the most metal poor systems simulated ($10^{-3}~Z_\odot$) the expected WR winds are still at least a factor of two larger than the line width observed in {\it Hebe}.

An additional problem with the WR scenario is that the fraction of WR stars decreases steeply at low metallicities \citep{Vink2005,Brinchmann2008} along with their line luminosities 
\citep{Crowther2006,Crowther2023}. This further suggests that WR stars are unlikely to contribute significantly at the extremely low metallicities.

Finally, the double peaked profile of the HeII emission in {\it Hebe} is not typical of WR stars, whose spectra are characterized by a Gaussian-like profile \citep[e.g.][]{Crowther2023}, and it is not clear why the HeII WR feature should have a velocity offset relative to the H$\gamma$.

\section{The direct collapse black hole or primordial black hole scenarios}
\label{sec:dcbh}

Another interesting possibility is that {\it Hebe} is actually a very small accreting black hole. If so, given the absence of metal lines (see \citealt{Rusta2026} and \citealt{Ubler2026} for metallicity constraints), this would be the most metal-poor accreting black hole ever discovered \citep[see ][for another extreme example]{Maiolino:2025}. Being in the vicinity of GN-z11, embedded in a strong Lyman-Werner radiation field, it would be the optimal environment for the formation of a Direct Collapse Black Hole \citep[DCBH, e.g.][]{Loeb1994,Bromm2003,Regan2017}. There is no evidence for a broad component of H$\gamma$ that would indicate the presence of a Broad Line Region. However, it could potentially be that the observed narrow H$\gamma$ is itself tracing the BLR of a very small black hole, with very low velocity dispersion. \citet{Ubler2026} infer an upper limit on the black hole mass of a few times $10^4~M_\odot$. This would be borderline with expectations for DCBHs, but still consistent with the predictions \citep{Ferrara2014}.

Another formation channel can be that of a primordial black hole \citep[PBH,][]{Hawking1971,Carr1974,Escriva2024}. These are expected to have a broad mass function, but recent constraints from the CMB suggest an upper limit of about $10^4~M_\odot$ \citep{Matteri2025}, which would more comfortably comply with the observational upper limit provided by \citet{Ubler2026}; although one should take into account the strong clustering expected for PBHs, which can result in a rapid growth via merging in the early Universe. In the PBH scenario, the vicinity of GN-z11 would allow the initially ``naked'' black holes to start accreting gas from the circumgalactic medium of the galaxy, as it plunges into its halo.

Other possible formation channels for massive black holes are stellar seeds accreting through super-Eddington bursts \citep[e.g.][]{Schneider2023,Trinca2023,Trinca2024, Banik2019}, or via rapid merging of stars and black holes in nuclear cluster \citep[e.g.][]{Partmann2025,Rantala2025}. However, in these cases, the star formation associated with the black hole seeding and subsequent accretion is expected to result in significant chemical enrichment. 

Regardless of the potential nature and origin,  an accreting black hole in such a metal-poor environment would be an extremely interesting finding. However, there are a few aspects that make these scenarios less plausible. First, the lower limits on the EW(HeII) are uncomfortably high for the AGN scenario, as illustrated in Fig.\ref{fig:diag}. Even stronger tension is found by \citet{Ubler2026} when exploring the H$\gamma$ emission, as they obtain a lower limit on EW(H$\gamma$) that is an order of magnitude higher than expected by accreting black hole models. Secondly, the observed HeII line profile is quite different from the H$\gamma$ one. While BLR stratification can broaden HeII with respect to Balmer lines \citep[e.g.][]{Brazzini2026}, HeII in AGNs is generally not strongly blueshifted, nor it presents a double-peaked profile, unless the Balmer lines also do.

\begin{figure}
	\centering
\includegraphics[width=\linewidth]{"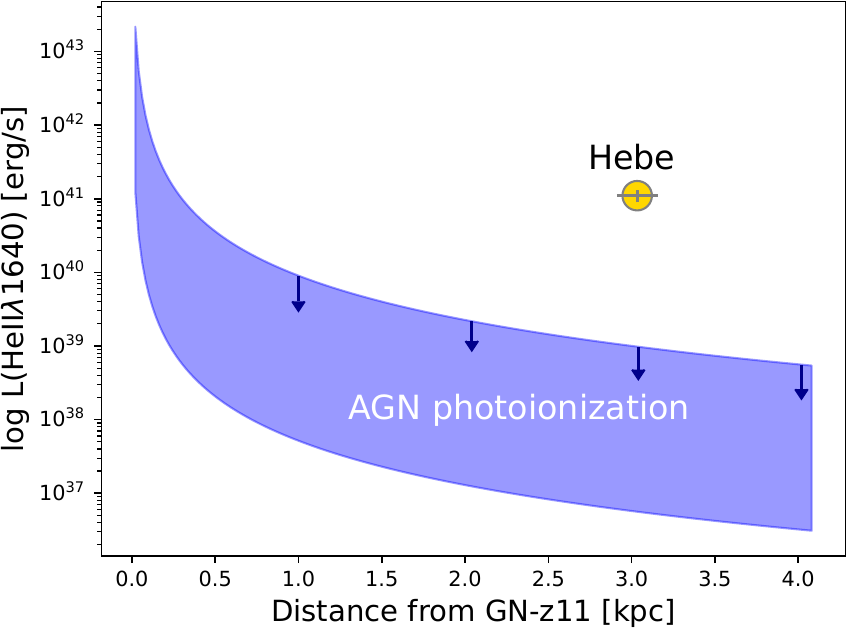"}
	\caption{Expected maximum HeII luminosity resulting from a cloud photoionized by the AGN in GN-z11, as a function of distance from the latter. These are conservative upper limits, as it is assumed that the projected distance of the putative cloud from GN-z11 is the actual distance (in addition to other conservative assumptions discussed in the text). The golden circle shows the HeII luminosity of {\it Hebe} at the observed projected distance. 
    }
\label{fig:heii_dist}
\end{figure}

\section{Excluding photoionization from the AGN in GN-z11}
\label{sec:heii_dist}

As mentioned in the introduction, \citet{Maiolino2024} identified a number of spectral features indicating the presence of a type 1 AGN in GN-z11. In particular, they found evidence of gas with densities $n_e>10^9~cm^{-3}$, typical of an AGN BLR, along with other AGN spectroscopic tracers, which were also identified in subsequent works  \citep[e.g.,][]{Ji2025GNz11, Scholtz2025_type2, CrespoGomez:2026, Fabian:2026}. 

{\it Hebe} could then be a pristine clump of gas photoionized by the AGN in GN-z11. This scenario can be tested by adopting (and updating) the same methodology used by \citet[][]{Maiolino:2024_Halo}, i.e. by estimating the HeII luminosity expected from a cloud illuminated by an AGN-like spectrum with the luminosity estimated in GN-z11 ($\sim 10^{45}$~erg/s), varying the distance of the cloud, and comparing it with the measured HeII line luminosity. This can be done under the conservative assumptions that the cloud has a size comparable with the size of the PSF (this is an upper limit, given that {\it Hebe} is unresolved), that it absorbs the entirety of the impinging ionizing photons from GN-z11 (while in reality the cloud could be clumpy and porous, or matter bounded, hence letting ionizing photons pass through), and that the projected distance is the actual cloud distance (in reality, it represents a lower limit).
Using the several AGN models in NM22, the blue shaded region in Fig.\ref{fig:heii_dist} shows the expected HeII luminosity of the ionized cloud as a function of distance from GN-z11 (the thickness of the region reflects the range of photoionization models), while the golden symbol shows the luminosity observed in {\it Hebe}.
Photoionization from the AGN in GN-z11 fails to account for the observed HeII emission by more than two orders of magnitude (given that these are conservative upper limits).

It may still be possible that the AGN in GN-z11 was far more luminous about $10^4$ years (rest frame) before the epoch of the observation (corresponding to the light-travel time to the cloud), and that the HeII emission is the fossil signature of such a previous activity. However, the previous AGN luminosity of GN-z11 should have been higher than $> 10^{47}$~erg/s, i.e. should have been one of the most luminous quasars in the whole universe \citep[e.g.,][]{Rakshit:2020}; yet, in contrast to the hyperluminous quasars, which are found at z$\sim 2-3$ when searching across the whole sky, GN-z11 should have been so hyperluminous already at z$>$10, and located in the narrow Chandra Deep Field North.

Yet another possibility is that the emission of GN-z11 is highly anisotropic. In this scenario,  {\it Hebe} could be seeing a radiation much larger than our line of sight. Specifically, this could be happening if GN-z11 is accreting at highly super-Eddington rate, in which case the ionizing radiation would be highly beamed in the direction perpendicular to the accretion disc \citep[e.g.][]{Madau2024}, and {\it Hebe} sees that disc nearly face on, while we are observing it close to edge on.
We have explored this scenario more quantitatively by leveraging the framework presented in \citet{Madau_Maiolino_2026} for Little Red Dots and Little Blue Dots. We obtain that even in this configuration, the model falls short in explaining the HeII luminosity of {\it Hebe} by at least an order of magnitude (especially given the very conservative assumptions on the distance and property of the putative clump). Additionally, the highly anisotropic and fortuitous orientation scenario, in such a small field, would imply that there should be many other similar AGN in the sky at $z>10$, pointing towards us with magnitude $<21$.

\section{Summary and Conclusions}
\label{sec:summary}

We have presented deep NIRSpec-IFU high-resolution (G235H) observations
of {\it Hebe}, a HeII emitting object in the halo of the z=10.6 galaxy GN-z11, previously identified via medium resolution (G235M) IFU spectroscopy.
The main results of the observations can be summarized as follows:

\begin{itemize}

\item HeII emission is confirmed by the G235H data, in a compact region, at the location of the previous G235M detection, 3~kpc from GN-z11.

\item The HeII identification is independently confirmed by the detection of H$\gamma$ in the same region by \citet{Ubler2026}.

\item No metal lines are detected.

\item The HeII emission is spectrally resolved in two components separated by about 120~km/s; the H$\gamma$ emission is primarily associated with one of them.

\item The EW(HeII) is extremely high, i.e. $>20$ \AA.

\end{itemize}

From these findings we infer the following:

\begin{itemize}

\item The only models consistent with the observed EW(HeII) and HeII/H$\gamma$ ratios are those involving PopIII stars. Further constraints supporting this scenario come from modelling the non-detection of metal lines, as discussed in a companion paper \citep{Rusta2026}.

\item We have explored the possibility that the HeII emission arises from a population of extremely metal poor WR stars. This scenario is very unlikely, based on the observed properties of GN-z11 and also based on the steeply decreasing abundance and luminosity of the WR stars at low metallicities.

\item We have also explored the possibility of
a Direct Collapse Black Hole or a Primordial Black Hole. However, these scenarios face significant issues. In particular, they struggle to reproduce the observed high EW(HeII) and EW(H$\gamma$), as well as the emission line profiles.

\item Photoionization by the AGN hosted in GN-z11 is safely ruled out.

In summary, {\it Hebe} represents one of the most convincing pieces of evidence for Population III stars in the early Universe, and its properties support models of their formation and early evolution.

\end{itemize}

\section*{Acknowledgments}

We are grateful to Richard Ellis, Jorryt Matthee, and Daniel Schaerer for useful comments.
This work is based on observations made with the National Aeronautics and Space Administration (NASA)/European Space Agency (ESA)/Canadian Space Agency (CSA) JWST. The data were obtained from the Mikulski Archive for Space Telescopes at the STScI, which is operated by the Association of Universities for Research in Astronomy, Inc., under NASA contract NAS 5-03127 for JWST. These observations are associated with programme PID 5086. RM, FD, JS, IJ, GJ acknowledge support from the Science and Technology Facilities Council (STFC), by the European Research Council (ERC) through Advanced Grant 695671 ``QUENCH'', by the UK Research and Innovation (UKRI) Frontier Research grant RISEandFALL. RM also acknowledges support from a Royal Society Research Professorship grant. AJB, GC and JC acknowledge funding from the “FirstGalaxies” Advanced Grant from the European Research Council (ERC) under the European Union’s Horizon 2020 research and innovation program (Grant agreement No. 789056).
H\"U thanks the Max Planck Society for support through the Lise Meitner Excellence Program. H\"U acknowledges funding by the European Union (ERC APEX, 101164796).
MP acknowledges support through the grants PID2021-127718NB-I00, PID2024-159902NA-I00, and RYC2023-044853-I, funded by the Spain Ministry of Science and Innovation/State Agency of Research MCIN/AEI/10.13039/501100011033 and El Fondo Social Europeo Plus FSE+.
JW gratefully acknowledges support from the Cosmic Dawn Center through the DAWN Fellowship. The Cosmic Dawn Center (DAWN) is funded by the Danish National Research Foundation under grant No. 140.
PGP-G acknowledges support from grant PID2022-139567NB-I00 funded by Spanish Ministerio de Ciencia, Innovaci\'on y Universidades MCIU/AEI/10.13039/501100011033,FEDER {\it Una manera de hacer Europa}.
JAAT acknowledges support from the Simons Foundation and \emph{JWST} program 3215. Support for program 3215 was provided by NASA through a grant from the Space Telescope Science Institute, which is operated by the Association of Universities for Research in Astronomy, Inc., under NASA contract NAS 5-03127.
BRP acknowledges support from
grants PID2021-127718NB-I00 and PID2024-158856NA-I00 funded by Spanish
Ministerio de Ciencia e Innovaci\'on y Universidades MCIU/AEI/10.13039/501100011033 and by
“ERDF A way of making Europe”. 
DJE acknowledges support by JWST/NIRCam contract to the University of Arizona, NAS5-02105, and as a Simons Foundation Investigator. EE, KH and CNAW acknowledge support by the JWST/NIRCam contract to the University of Arizona NAS5-0215.
SC, GV, and SZ acknowledge support by European Union’s HE ERC Starting Grant No. 101040227 - WINGS.
WMB gratefully acknowledges support from DARK via the DARK fellowship. This work was supported by a research grant (VIL54489) from VILLUM FONDEN.
JMH acknowledges support from JWST Program \#8544.
Views and opinions expressed are those of the authors only and do not necessarily reflect those of the European Union or the European Research Council Executive Agency. Neither the European Union nor the granting authority can be held responsible for them.

\section*{Data availability}

The NIRSpec data used in this study are publicly available at the STScI MAST archive: \url{https://mast.stsci.edu/portal/Mashup/Clients/Mast/Portal.html}, under GO programme 5086.

\appendix

\section{Single Gaussian simultaneous HeII and \texorpdfstring{H$\gamma$}{Hg} fit}
\label{app:oneGfit}

\begin{figure}
	\centering
    \includegraphics[width=0.9\linewidth]{"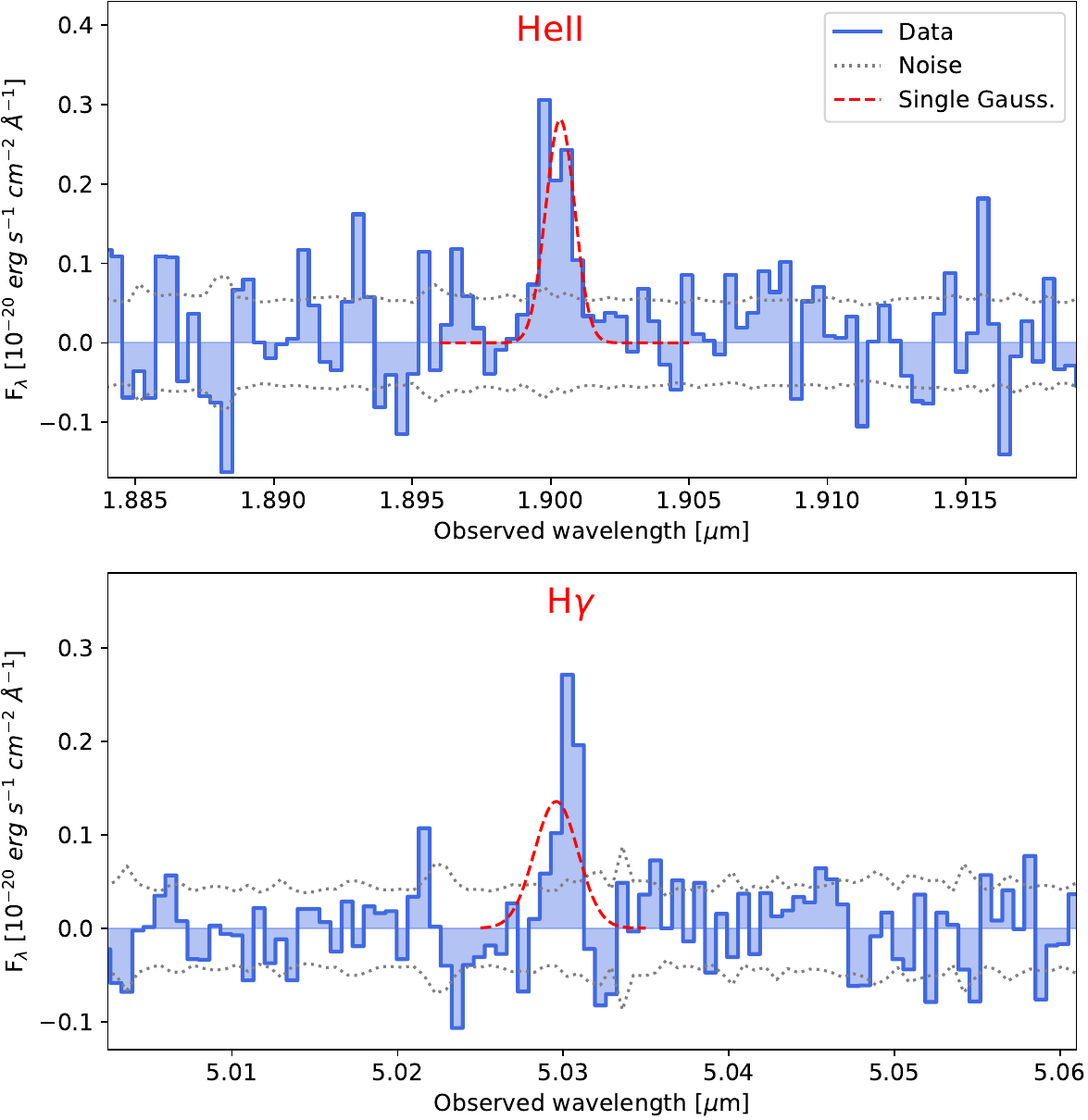"}
	\caption{HeII and H$\gamma$ spectra of {\it Hebe} fitted simultaneously with a single Gaussian (i.e. the same velocity and width for the Gaussians ifitting the two lines).
	}
	\label{fig:oneGfit}
\end{figure}

Fig.\ref{fig:oneGfit} shows the result of simultaneously fitting HeII and H$\gamma$ with a single Gaussian, i.e. the the two lines are fitted with Gaussians with the same velocity and width. 
In this case $BIC_{1G}=87$, in contrast with the case of fitting with two Gaussian components ($BIC_{2G}=65$), resulting in a $\Delta BIC=22$ -- a clear indication that the fit with a single Gaussian fit is much worse than the fit with two Gaussian components.

\section{Metal lines non detection}
\label{app:metallines}

\begin{figure}
	\centering
    \includegraphics[width=0.9\linewidth]{"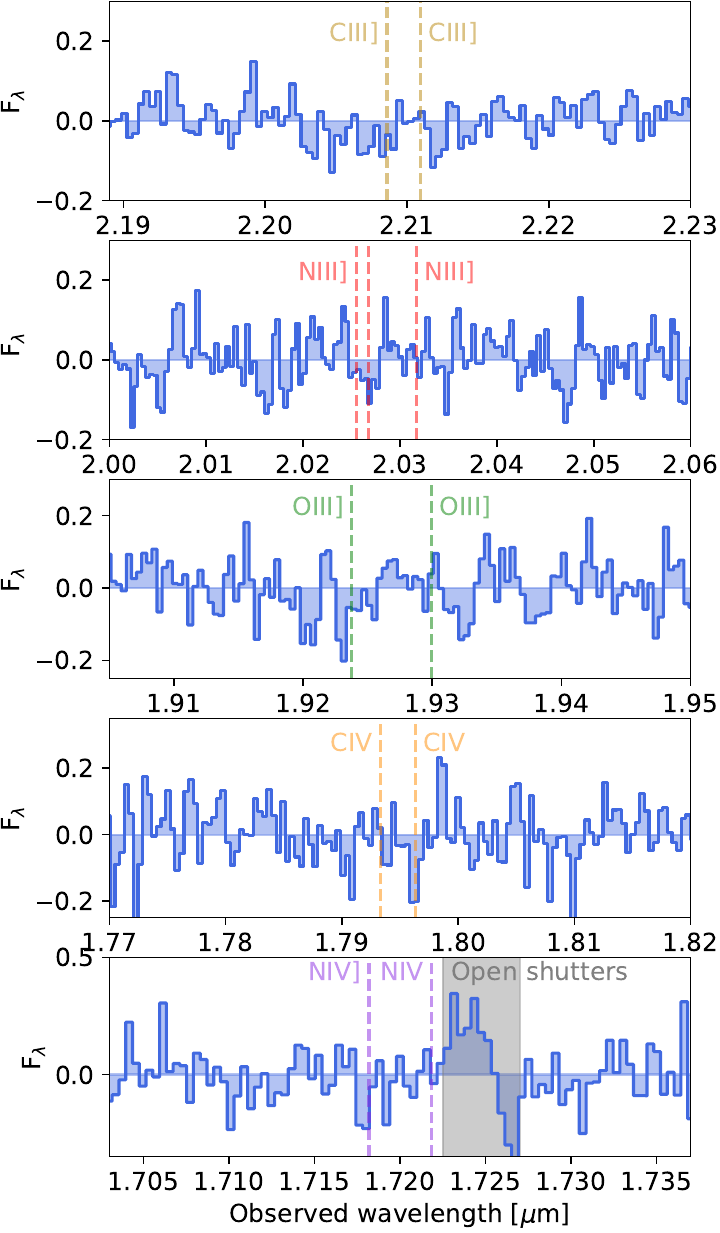"}
	\caption{Spectrum of {\it Hebe} zoomed at the wavelengths where typically the strongest metal lines are observed in normal galaxies, specifically, from top to bottom: CIII]1906,1908, OIII]1661,1666, NIII]1749 multiplet, CIV1548,1551, and NIV]1483,1486.
    The gray shaded region in the bottom panel marks an unmasked faint open shutter/artefact affecting that spectral region.
    The units of the y axis are in $10^{-20} ~\rm erg~s^{-1}~cm^{-2}$~\AA$^{-1}$.
	}
	\label{fig:metallines}
\end{figure}

Fig.\ref{fig:metallines} shows the spectrum of {\it Hebe} zoomed at the wavelengths where typically the strongest metal lines are observed in normal galaxies, specifically, from top to bottom: CIII]1906,1908, OIII]1661,1666, NIII]1749 multiplet, CIV1548,1551, and NIV]1483,1486. Note that an unmasked faint open shutter/artefact affects a region near the NIV doublet (gray shaded region).

\section{Gravitational lens modeling}\label{lens_app}
Both GNz-11 and {\it Hebe} are in close proximity (i.e., within $\sim1$ projected arcsecond) of a foreground galaxy at $z=2.028$. In order to determine the lensing magnification of each object, we use the code \textlcsc{PyAutoLens} \citep{pyautolens} to construct the gravitational lens model of this field. 

First, we collapse our data cube around the redshifted wavelength of H$\alpha$ of the foreground galaxy ($1.9851<\lambda_{\rm obs}/\mu \rm m<1.9905$), resulting in a map containing a strong, resolved detection of the foreground galaxy as well as a continuum detection of GNz-11. We then create a model of this field by using S\'ersic light profiles for the two galaxies, and use \textlcsc{PyAutoLens} to determine the best-fit properties of each. The mass of the foreground galaxy is assumed to be an isothermal profile with the same spatial centroid, axis ratio, and position angle as the best-fit S\'ersic profile. The Einstein radius of this mass profile is determined assuming a total (i.e., baryonic and non-baryonic) mass of $\sim10^{10.4}\,M_{\odot}$, as derived through kinematic modeling of this source. 

This approach results in gravitational magnification estimates of $\mu=1.4$ for {\it Hebe} and $\mu=1.1$ for GNz-11. These values are dependent on the total mass of the foreground galaxy, so there is some uncertainty in each. If we instead adopt the conservative total mass limit of \citet[][$\sim10^{10.9}\,M_{\odot}$]{Tacchella2023}, which was determined for the total halo mass containing GNz-11, then we find slightly higher magnification for both {\it Hebe} ($\mu\sim2.5$) and GNz-11 ($\mu\sim1.3$). If we instead adopt the best-fit stellar mass based on SED fits ($\sim10^{8.9}\,M_{\odot}$), then we find that each galaxy is only magnified by a few percent. In any case, we may rule out very high magnification factors of $\mu>10$ for each source.



\bibliographystyle{mnras}
\bibliography{roberto} 

@ARTICLE{Abel+2002,
       author = {{Abel}, Tom and {Bryan}, Greg L. and {Norman}, Michael L.},
        title = "{The Formation of the First Star in the Universe}",
      journal = {Science},
         year = 2002,
        month = jan,
       volume = {295},
       number = {5552},
        pages = {93-98},
          doi = {10.1126/science.1063991},
archivePrefix = {arXiv},
       eprint = {astro-ph/0112088},
 primaryClass = {astro-ph},
       adsurl = {https://ui.adsabs.harvard.edu/abs/2002Sci...295...93A}
}

@ARTICLE{Yoshida+2003,
       author = {{Yoshida}, Naoki and {Abel}, Tom and {Hernquist}, Lars and {Sugiyama}, Naoshi},
        title = "{Simulations of Early Structure Formation: Primordial Gas Clouds}",
      journal = {\apj},
         year = 2003,
        month = aug,
       volume = {592},
       number = {2},
        pages = {645-663},
          doi = {10.1086/375810},
archivePrefix = {arXiv},
       eprint = {astro-ph/0301645},
 primaryClass = {astro-ph},
       adsurl = {https://ui.adsabs.harvard.edu/abs/2003ApJ...592..645Y}
}

@ARTICLE{Sun2026,
       author = {{Sun}, Fengwu and {Eisenstein}, Daniel J. and {D'Eugenio}, Francesco and {Hainline}, Kevin and {Helton}, Jakob M. and {Johnson}, Benjamin D. and {Lin}, Xiaojing and {Rieke}, Marcia and {Robertson}, Brant and {Tacchella}, Sandro and {Bunker}, Andrew J. and {Chevallard}, Jacopo and {Curtis-Lake}, Emma and {Egami}, Eiichi and {Hausen}, Ryan and {Ji}, Zhiyuan and {Lyu}, Jianwei and {Maiolino}, Roberto and {Rinaldi}, Pierluigi and {Sun}, Yang and {Trussler}, James A.~A. and {Williams}, Christina C. and {Willmer}, Christopher N.~A. and {Witstok}, Joris and {Wu}, Zihao and {Zhu}, Yongda},
        title = "{JADES: Discovery of Large Reservoirs of Small Dust Grains in the Circumgalactic Medium of Massive Galaxies at $z\sim3.5$ through Deep JWST/NIRCam Imaging and Grism Spectroscopy}",
      journal = {arXiv e-prints},
         year = 2026,
        month = jan,
          eid = {arXiv:2601.15961},
        pages = {arXiv:2601.15961},
          doi = {10.48550/arXiv.2601.15961},
archivePrefix = {arXiv},
       eprint = {2601.15961},
 primaryClass = {astro-ph.GA},
       adsurl = {https://ui.adsabs.harvard.edu/abs/2026arXiv260115961S}
}

@ARTICLE{Sibony2022,
       author = {{Sibony}, Y. and {Liu}, B. and {Simmonds}, C. and {Meynet}, G. and {Bromm}, V.},
        title = "{Impact of Population III homogeneous stellar evolution on early cosmic reionisation}",
      journal = {\aap},
         year = 2022,
        month = oct,
       volume = {666},
          eid = {A199},
        pages = {A199},
          doi = {10.1051/0004-6361/202244146},
archivePrefix = {arXiv},
       eprint = {2205.15125},
 primaryClass = {astro-ph.SR},
       adsurl = {https://ui.adsabs.harvard.edu/abs/2022A&A...666A.199S}
}

@ARTICLE{Murphy2021,
       author = {{Murphy}, Laura J. and {Groh}, Jose H. and {Farrell}, Eoin and {Meynet}, Georges and {Ekstr{\"o}m}, Sylvia and {Tsiatsiou}, Sophie and {Hackett}, Alexander and {Martinet}, S{\'e}bastien},
        title = "{Ionizing photon production of Population III stars: effects of rotation, convection, and initial mass function}",
      journal = {\mnras},
         year = 2021,
        month = oct,
       volume = {506},
       number = {4},
        pages = {5731-5749},
          doi = {10.1093/mnras/stab2073},
archivePrefix = {arXiv},
       eprint = {2105.06900},
 primaryClass = {astro-ph.SR},
       adsurl = {https://ui.adsabs.harvard.edu/abs/2021MNRAS.506.5731M}
}

@ARTICLE{Yoon2012,
       author = {{Yoon}, S.-C. and {Dierks}, A. and {Langer}, N.},
        title = "{Evolution of massive Population III stars with rotation and magnetic fields}",
      journal = {\aap},
         year = 2012,
        month = jun,
       volume = {542},
          eid = {A113},
        pages = {A113},
          doi = {10.1051/0004-6361/201117769},
archivePrefix = {arXiv},
       eprint = {1201.2364},
 primaryClass = {astro-ph.SR},
       adsurl = {https://ui.adsabs.harvard.edu/abs/2012A&A...542A.113Y}
}

@ARTICLE{Wasserman2026,
       author = {{Wasserman}, Joel and {Zackrisson}, Erik and {Dhandha}, Jiten and {Fialkov}, Anastasia and {Noble}, Leon and {Majumdar}, Suman},
        title = "{Ultraviolet photon production rates of the first stars: Impact on the He II {\ensuremath{\lambda}} 1640 {\r{A}} emission line from primordial star clusters and the 21-cm signal from cosmic dawn}",
      journal = {\mnras},
         year = 2026,
        month = feb,
          doi = {10.1093/mnras/stag386},
archivePrefix = {arXiv},
       eprint = {2507.21764},
 primaryClass = {astro-ph.GA},
       adsurl = {https://ui.adsabs.harvard.edu/abs/2026MNRAS.tmp..379W}
}

@ARTICLE{Martinez2025,
       author = {{Martinez}, Zorayda and {Berg}, Danielle A. and {James}, Bethan L. and {Arellano-C{\'o}rdova}, Karla Z. and {Stark}, Daniel P. and {Senchyna}, Peter and {Skillman}, Evan D. and {Rogers}, Noah S.~J. and {Chisholm}, John},
        title = "{Under Pressure: Decoding the Effect of High Densities on Derived Nebular Properties}",
      journal = {\apj},
         year = 2025,
        month = dec,
       volume = {995},
       number = {2},
          eid = {204},
        pages = {204},
          doi = {10.3847/1538-4357/ae17c6},
archivePrefix = {arXiv},
       eprint = {2510.21960},
 primaryClass = {astro-ph.GA},
       adsurl = {https://ui.adsabs.harvard.edu/abs/2025ApJ...995..204M}
}

@ARTICLE{Scholtz2025DR4,
       author = {{Scholtz}, J. and {Carniani}, S. and {Parlanti}, E. and {D'Eugenio}, F. and {Curtis-Lake}, E. and {Jakobsen}, P. and {Bunker}, A.~J. and {Cameron}, A.~J. and {Arribas}, S. and {Baker}, W.~M. and {Charlot}, S. and {Chevellard}, J. and {Circosta}, C. and {Curti}, M. and {Duan}, Q. and {Eisenstein}, D.~J. and {Hainline}, K. and {Ji}, Z. and {Johnson}, B.~D. and {Jones}, G.~C. and {Kumari}, N. and {Maiolino}, R. and {Maseda}, M.~V. and {Perna}, M. and {P{\'e}rez-Gonz{\'a}lez}, P.~G. and {Rawle}, T. and {Rieke}, M. and {Rinaldi}, P. and {Robertson}, B. and {Saxena}, A. and {Shivaei}, I. and {Silcock}, M.~S. and {Sun}, Y. and {Rodr{\'\i}guez Del Pino}, B. and {Tacchella}, S. and {{\"U}bler}, H. and {Venturi}, G. and {Williams}, C.~C. and {Willmer}, C.~N.~A. and {Willott}, C. and {Witstok}, J.},
        title = "{JADES Data Release 4 -- Paper II: Data reduction, analysis and emission-line fluxes of the complete spectroscopic sample}",
      journal = {arXiv e-prints},
         year = 2025,
        month = oct,
          eid = {arXiv:2510.01034},
        pages = {arXiv:2510.01034},
          doi = {10.48550/arXiv.2510.01034},
archivePrefix = {arXiv},
       eprint = {2510.01034},
 primaryClass = {astro-ph.GA},
       adsurl = {https://ui.adsabs.harvard.edu/abs/2025arXiv251001034S}
}

@ARTICLE{boeker2022,
       author = {{B{\"o}ker}, T. and {Arribas}, S. and {L{\"u}tzgendorf}, N. and {Alves de Oliveira}, C. and {Beck}, T.~L. and {Birkmann}, S. and {Bunker}, A.~J. and {Charlot}, S. and {de Marchi}, G. and {Ferruit}, P. and {Giardino}, G. and {Jakobsen}, P. and {Kumari}, N. and {L{\'o}pez-Caniego}, M. and {Maiolino}, R. and {Manjavacas}, E. and {Marston}, A. and {Moseley}, S.~H. and {Muzerolle}, J. and {Ogle}, P. and {Pirzkal}, N. and {Rauscher}, B. and {Rawle}, T. and {Rix}, H.-W. and {Sabbi}, E. and {Sargent}, B. and {Sirianni}, M. and {te Plate}, M. and {Valenti}, J. and {Willott}, C.~J. and {Zeidler}, P.},
        title = "{The Near-Infrared Spectrograph (NIRSpec) on the James Webb Space Telescope. III. Integral-field spectroscopy}",
      journal = {\aap},
         year = 2022,
        month = may,
       volume = {661},
          eid = {A82},
        pages = {A82},
          doi = {10.1051/0004-6361/202142589},
archivePrefix = {arXiv},
       eprint = {2202.03308},
 primaryClass = {astro-ph.IM},
       adsurl = {https://ui.adsabs.harvard.edu/abs/2022A&A...661A..82B}
}

@ARTICLE{Wechsler2018,
       author = {{Wechsler}, Risa H. and {Tinker}, Jeremy L.},
        title = "{The Connection Between Galaxies and Their Dark Matter Halos}",
      journal = {\araa},
         year = 2018,
        month = sep,
       volume = {56},
        pages = {435-487},
          doi = {10.1146/annurev-astro-081817-051756},
archivePrefix = {arXiv},
       eprint = {1804.03097},
 primaryClass = {astro-ph.GA},
       adsurl = {https://ui.adsabs.harvard.edu/abs/2018ARA&A..56..435W}
}

@ARTICLE{Schaerer2003,
       author = {{Schaerer}, D.},
        title = "{The transition from Population III to normal galaxies: Lyalpha and He II emission and the ionising properties of high redshift starburst galaxies}",
      journal = {\aap},
         year = 2003,
        month = jan,
       volume = {397},
        pages = {527-538},
          doi = {10.1051/0004-6361:20021525},
archivePrefix = {arXiv},
       eprint = {astro-ph/0210462},
 primaryClass = {astro-ph},
       adsurl = {https://ui.adsabs.harvard.edu/abs/2003A&A...397..527S}
}

@ARTICLE{Zackrisson2011,
       author = {{Zackrisson}, Erik and {Rydberg}, Claes-Erik and {Schaerer}, Daniel and {{\"O}stlin}, G{\"o}ran and {Tuli}, Manan},
        title = "{The Spectral Evolution of the First Galaxies. I. James Webb Space Telescope Detection Limits and Color Criteria for Population III Galaxies}",
      journal = {\apj},
         year = 2011,
        month = oct,
       volume = {740},
       number = {1},
          eid = {13},
        pages = {13},
          doi = {10.1088/0004-637X/740/1/13},
archivePrefix = {arXiv},
       eprint = {1105.0921},
 primaryClass = {astro-ph.CO},
       adsurl = {https://ui.adsabs.harvard.edu/abs/2011ApJ...740...13Z}
}

@ARTICLE{Bromm2002,
       author = {{Bromm}, Volker and {Coppi}, Paolo S. and {Larson}, Richard B.},
        title = "{The Formation of the First Stars. I. The Primordial Star-forming Cloud}",
      journal = {\apj},
         year = 2002,
        month = jan,
       volume = {564},
       number = {1},
        pages = {23-51},
          doi = {10.1086/323947},
archivePrefix = {arXiv},
       eprint = {astro-ph/0102503},
 primaryClass = {astro-ph},
       adsurl = {https://ui.adsabs.harvard.edu/abs/2002ApJ...564...23B}
}

@ARTICLE{Schneider2002,
       author = {{Schneider}, R. and {Ferrara}, A. and {Natarajan}, P. and {Omukai}, K.},
        title = "{First Stars, Very Massive Black Holes, and Metals}",
      journal = {\apj},
         year = 2002,
        month = may,
       volume = {571},
       number = {1},
        pages = {30-39},
          doi = {10.1086/339917},
archivePrefix = {arXiv},
       eprint = {astro-ph/0111341},
 primaryClass = {astro-ph},
       adsurl = {https://ui.adsabs.harvard.edu/abs/2002ApJ...571...30S}
}

@ARTICLE{Berg2025,
       author = {{Berg}, Danielle A. and {Naidu}, Rohan P. and {Chisholm}, John and {Atek}, Hakim and {Fujimoto}, Seiji and {Kokorev}, Vasily and {Furtak}, Lukas J. and {Kobayashi}, Chiaki and {Schaerer}, Daniel and {Adamo}, Angela and {Fei}, Qinyue and {Korber}, Damien and {Matthee}, Jorryt and {Marques-Chaves}, Rui and {Martinez}, Zorayda and {Mcquinn}, Kristen B.~W. and {Mu{\~n}oz}, Julian B. and {Oesch}, Pascal A. and {Stark}, Daniel P. and {Stephenson}, Mabel G. and {Hsiao}, Tiger Yu-Yang},
        title = "{A Fleeting GLIMPSE of N/O Enrichment at Cosmic Dawn: Evidence for Wolf Rayet N Stars in a z = 6.1 Galaxy}",
      journal = {arXiv e-prints},
         year = 2025,
        month = nov,
          eid = {arXiv:2511.13591},
        pages = {arXiv:2511.13591},
          doi = {10.48550/arXiv.2511.13591},
archivePrefix = {arXiv},
       eprint = {2511.13591},
 primaryClass = {astro-ph.GA},
       adsurl = {https://ui.adsabs.harvard.edu/abs/2025arXiv251113591B}
}

@ARTICLE{Ubler2026,
       author = {{{\"U}bler}, Hannah and {Maiolino}, Roberto and {P{\'e}rez-Gonz{\'a}lez}, Pablo G. and {Isobe}, Yuki and {Jones}, Gareth C. and {Kumari}, Nimisha and {Charlot}, St{\'e}phane and {Rusta}, Elka and {Salvadori}, Stefania and {Nakajima}, Kimihiko and {Perna}, Michele and {Arribas}, Santiago and {Bunker}, Andrew J. and {Carniani}, Stefano and {D'Eugenio}, Francesco and {Rodr{\'\i}guez Del Pino}, Bruno and {Bertola}, Elena and {B{\"o}ker}, Torsten and {Chevallard}, Jacopo and {Circosta}, Chiara and {Cresci}, Giovanni and {Curti}, Mirko and {Curtis-Lake}, Emma and {Eisenstein}, Daniel J. and {Hainline}, Kevin and {Johnson}, Benjamin D. and {Parlanti}, Eleonora and {Rinaldi}, Pierluigi and {Robertson}, Brant and {Scholtz}, Jan and {Tacchella}, Sandro and {Venturi}, Giacomo and {Witstok}, Joris and {Zamora}, Sandra},
        title = "{GA-NIFS \& JADES: Confirmation of pristine gas near GN-z11}",
      journal = {arXiv e-prints},
         year = 2026,
        month = mar,
          eid = {arXiv:2603.20360},
        pages = {arXiv:2603.20360},
archivePrefix = {arXiv},
       eprint = {2603.20360},
 primaryClass = {astro-ph.GA},
       adsurl = {https://ui.adsabs.harvard.edu/abs/2026arXiv260320360U}
}

@ARTICLE{Rusta2026,
       author = {{Rusta}, Elka and {Salvadori}, Stefania and {Maiolino}, Roberto and {Gelli}, Viola and {Koutsouridou}, Ioanna and {Carniani}, Stefano and {{\"U}bler}, Hannah and {Marconi}, Alessandro and {Schaerer}, Daniel},
        title = "{The Pristine HeII Emitter near GN-z11: Constraining the Mass Distribution of the First Stars}",
      journal = {arXiv e-prints},
         year = 2026,
        month = mar,
          eid = {arXiv:2603.20363},
        pages = {arXiv:2603.20363},
archivePrefix = {arXiv},
       eprint = {2603.20363},
 primaryClass = {astro-ph.GA},
       adsurl = {https://ui.adsabs.harvard.edu/abs/2026arXiv260320363R}
}

@ARTICLE{Moreschini2026,
       author = {{Moreschini}, Bianca and {Belfiore}, Francesco and {Marconi}, Alessandro and {Cataldi}, Elisa and {Curti}, Mirko and {Amiri}, Amirnezam and {Feltre}, Anna and {Mannucci}, Filippo and {Bertola}, Elena and {Bracci}, Caterina and et al.},
        title = "{One cloud is not enough: extreme conditions bias chemical abundances in high-redshift galaxies}",
      journal = {arXiv e-prints},
         year = 2026,
        month = jan,
          eid = {arXiv:2601.08939},
        pages = {arXiv:2601.08939},
          doi = {10.48550/arXiv.2601.08939},
archivePrefix = {arXiv},
       eprint = {2601.08939},
 primaryClass = {astro-ph.GA},
       adsurl = {https://ui.adsabs.harvard.edu/abs/2026arXiv260108939M}
}

@ARTICLE{Banik2019,
       author = {{Banik}, Nilanjan and {Tan}, Jonathan C. and {Monaco}, Pierluigi},
        title = "{The formation of supermassive black holes from Population III.1 seeds. I. Cosmic formation histories and clustering properties}",
      journal = {\mnras},
         year = 2019,
        month = mar,
       volume = {483},
       number = {3},
        pages = {3592-3606},
          doi = {10.1093/mnras/sty3298},
archivePrefix = {arXiv},
       eprint = {1608.04421},
 primaryClass = {astro-ph.GA},
       adsurl = {https://ui.adsabs.harvard.edu/abs/2019MNRAS.483.3592B}
}

@ARTICLE{Berg2024,
       author = {{Berg}, Danielle A. and {Skillman}, Evan D. and {Chisholm}, John and {Pogge}, Richard W. and {Gazagnes}, Simon and {Rogers}, Noah S.~J. and {Erb}, Dawn K. and {Arellano-C{\'o}rdova}, Karla Z. and {Leitherer}, Claus and {Appel}, Jackie and {Moustakas}, John},
        title = "{CHAOS. VIII. Far-ultraviolet Spectra of M101 and the Impact of Wolf{\textendash}Rayet Stars}",
      journal = {\apj},
         year = 2024,
        month = aug,
       volume = {971},
       number = {1},
          eid = {87},
        pages = {87},
          doi = {10.3847/1538-4357/ad5292},
archivePrefix = {arXiv},
       eprint = {2405.19477},
 primaryClass = {astro-ph.GA},
       adsurl = {https://ui.adsabs.harvard.edu/abs/2024ApJ...971...87B}
}

@ARTICLE{Lecroq2024,
       author = {{Lecroq}, Marie and {Charlot}, St{\'e}phane and {Bressan}, Alessandro and {Bruzual}, Gustavo and {Costa}, Guglielmo and {Iorio}, Giuliano and {Spera}, Mario and {Mapelli}, Michela and {Chen}, Yang and {Chevallard}, Jacopo and {Dall'Amico}, Marco},
        title = "{Nebular emission from young stellar populations including binary stars}",
      journal = {\mnras},
         year = 2024,
        month = jan,
       volume = {527},
       number = {3},
        pages = {9480-9504},
          doi = {10.1093/mnras/stad3838},
archivePrefix = {arXiv},
       eprint = {2312.08432},
 primaryClass = {astro-ph.GA},
       adsurl = {https://ui.adsabs.harvard.edu/abs/2024MNRAS.527.9480L}
}

@ARTICLE{Brazzini2026,
       author = {{Brazzini}, M. and {D'Eugenio}, F. and {Maiolino}, R. and {Lyu}, J. and {DeCoursey}, C. and {{\"U}bler}, H. and {Ji}, X. and {Juod{\v{z}}balis}, I. and {Scholtz}, J. and {Jones}, G.~C. and {Hainline}, K. and {Dalla Bont{\`a}}, E. and {{\'e}rez-Gonz{\'a}lez}, P.~G. P and {Geris}, S. and {Harshan}, A. and {Feruglio}, C. and {Bischetti}, M. and {Mazzolari}, G. and {Rieke}, G. and {Alberts}, S. and {Trefoloni}, B. and {Carniani}, S. and {Parlanti}, E. and {Marconi}, A. and {Risaliti}, G. and {Ramos Almeida}, C. and {Rinaldi}, P. and {Perna}, M. and {Zamora}, S. and {Lamperti}, I. and {Venturi}, G. and {Cresci}, G. and {Bunker}, Andrew J. and {Ivey}, L.~R.},
        title = "{The Little Blue and Red Dots Rosetta Stones: Non-Gaussian broad lines, hot dust, and X-ray weakness}",
      journal = {arXiv e-prints},
         year = 2026,
        month = jan,
          eid = {arXiv:2601.22214},
        pages = {arXiv:2601.22214},
          doi = {10.48550/arXiv.2601.22214},
archivePrefix = {arXiv},
       eprint = {2601.22214},
 primaryClass = {astro-ph.GA},
       adsurl = {https://ui.adsabs.harvard.edu/abs/2026arXiv260122214B}
}

@ARTICLE{Matteri2025,
       author = {{Matteri}, Antonio and {Pallottini}, Andrea and {Ferrara}, Andrea},
        title = "{Can primordial black holes explain the overabundance of bright super-early galaxies?}",
      journal = {\aap},
         year = 2025,
        month = may,
       volume = {697},
          eid = {A65},
        pages = {A65},
          doi = {10.1051/0004-6361/202553701},
archivePrefix = {arXiv},
       eprint = {2503.01968},
 primaryClass = {astro-ph.GA},
       adsurl = {https://ui.adsabs.harvard.edu/abs/2025A&A...697A..65M}
}

@ARTICLE{Ferrara2014,
       author = {{Ferrara}, A. and {Salvadori}, S. and {Yue}, B. and {Schleicher}, D.},
        title = "{Initial mass function of intermediate-mass black hole seeds}",
      journal = {\mnras},
         year = 2014,
        month = sep,
       volume = {443},
       number = {3},
        pages = {2410-2425},
          doi = {10.1093/mnras/stu1280},
archivePrefix = {arXiv},
       eprint = {1406.6685},
 primaryClass = {astro-ph.GA},
       adsurl = {https://ui.adsabs.harvard.edu/abs/2014MNRAS.443.2410F},
}

@ARTICLE{Carr1974,
       author = {{Carr}, B.~J. and {Hawking}, S.~W.},
        title = "{Black holes in the early Universe}",
      journal = {\mnras},
         year = 1974,
        month = aug,
       volume = {168},
        pages = {399-416},
          doi = {10.1093/mnras/168.2.399},
       adsurl = {https://ui.adsabs.harvard.edu/abs/1974MNRAS.168..399C}
}

@ARTICLE{Hawking1971,
       author = {{Hawking}, Stephen},
        title = "{Gravitationally collapsed objects of very low mass}",
      journal = {\mnras},
         year = 1971,
        month = jan,
       volume = {152},
        pages = {75},
          doi = {10.1093/mnras/152.1.75},
       adsurl = {https://ui.adsabs.harvard.edu/abs/1971MNRAS.152...75H}
}

@ARTICLE{Regan2017,
       author = {{Regan}, John A. and {Visbal}, Eli and {Wise}, John H. and {Haiman}, Zolt{\'a}n and {Johansson}, Peter H. and {Bryan}, Greg L.},
        title = "{Rapid formation of massive black holes in close proximity to embryonic protogalaxies}",
      journal = {Nature Astronomy},
         year = 2017,
        month = mar,
       volume = {1},
          eid = {0075},
        pages = {0075},
          doi = {10.1038/s41550-017-0075},
archivePrefix = {arXiv},
       eprint = {1703.03805},
 primaryClass = {astro-ph.GA},
       adsurl = {https://ui.adsabs.harvard.edu/abs/2017NatAs...1E..75R}
}

@ARTICLE{Bromm2003,
       author = {{Bromm}, Volker and {Loeb}, Abraham},
        title = "{Formation of the First Supermassive Black Holes}",
      journal = {\apj},
         year = 2003,
        month = oct,
       volume = {596},
       number = {1},
        pages = {34-46},
          doi = {10.1086/377529},
archivePrefix = {arXiv},
       eprint = {astro-ph/0212400},
 primaryClass = {astro-ph},
       adsurl = {https://ui.adsabs.harvard.edu/abs/2003ApJ...596...34B}
}

@ARTICLE{Loeb1994,
       author = {{Loeb}, Abraham and {Rasio}, Frederic A.},
        title = "{Collapse of Primordial Gas Clouds and the Formation of Quasar Black Holes}",
      journal = {\apj},
         year = 1994,
        month = sep,
       volume = {432},
        pages = {52},
          doi = {10.1086/174548},
archivePrefix = {arXiv},
       eprint = {astro-ph/9401026},
 primaryClass = {astro-ph},
       adsurl = {https://ui.adsabs.harvard.edu/abs/1994ApJ...432...52L}
}

@ARTICLE{Brinchmann2008,
       author = {{Brinchmann}, J. and {Kunth}, D. and {Durret}, F.},
        title = "{Galaxies with Wolf-Rayet signatures in the low-redshift Universe. A survey using the Sloan Digital Sky Survey}",
      journal = {\aap},
         year = 2008,
        month = jul,
       volume = {485},
       number = {3},
        pages = {657-677},
          doi = {10.1051/0004-6361:200809783},
archivePrefix = {arXiv},
       eprint = {0805.1073},
 primaryClass = {astro-ph},
       adsurl = {https://ui.adsabs.harvard.edu/abs/2008A&A...485..657B}
}

@ARTICLE{Vink2005,
       author = {{Vink}, Jorick S. and {de Koter}, A.},
        title = "{On the metallicity dependence of Wolf-Rayet winds}",
      journal = {\aap},
         year = 2005,
        month = nov,
       volume = {442},
       number = {2},
        pages = {587-596},
          doi = {10.1051/0004-6361:20052862},
archivePrefix = {arXiv},
       eprint = {astro-ph/0507352},
 primaryClass = {astro-ph},
       adsurl = {https://ui.adsabs.harvard.edu/abs/2005A&A...442..587V}
}

@ARTICLE{Boco2025,
       author = {{Boco}, Lumen and {Mapelli}, Michela and {Sander}, Andreas A.~C. and {Mesini}, Sofia and {Ramachandran}, Varsha and {Torniamenti}, Stefano and {Korb}, Erika and {Liu}, Boyuan and {Sabhahit}, Gautham N. and {Vink}, Jorick S.},
        title = "{Metal-poor single Wolf-Rayet stars: The interplay of optically thick winds and rotation}",
      journal = {\aap},
         year = 2025,
        month = nov,
       volume = {703},
          eid = {A243},
        pages = {A243},
          doi = {10.1051/0004-6361/202556187},
archivePrefix = {arXiv},
       eprint = {2507.00137},
 primaryClass = {astro-ph.SR},
       adsurl = {https://ui.adsabs.harvard.edu/abs/2025A&A...703A.243B}
}

@ARTICLE{Gafener2008,
       author = {{Gr{\"a}fener}, G. and {Hamann}, W.-R.},
        title = "{Mass loss from late-type WN stars and its Z-dependence. Very massive stars approaching the Eddington limit}",
      journal = {\aap},
         year = 2008,
        month = may,
       volume = {482},
       number = {3},
        pages = {945-960},
          doi = {10.1051/0004-6361:20066176},
archivePrefix = {arXiv},
       eprint = {0803.0866},
 primaryClass = {astro-ph},
       adsurl = {https://ui.adsabs.harvard.edu/abs/2008A&A...482..945G}
}

@ARTICLE{Sabhahit2023,
       author = {{Sabhahit}, Gautham N. and {Vink}, Jorick S. and {Sander}, Andreas A.~C. and {Higgins}, Erin R.},
        title = "{Very massive stars and pair-instability supernovae: mass-loss framework for low metallicity}",
      journal = {\mnras},
         year = 2023,
        month = sep,
       volume = {524},
       number = {1},
        pages = {1529-1546},
          doi = {10.1093/mnras/stad1888},
archivePrefix = {arXiv},
       eprint = {2306.11785},
 primaryClass = {astro-ph.SR},
       adsurl = {https://ui.adsabs.harvard.edu/abs/2023MNRAS.524.1529S}
}

@ARTICLE{Hainich2014,
       author = {{Hainich}, R. and {R{\"u}hling}, U. and {Todt}, H. and {Oskinova}, L.~M. and {Liermann}, A. and {Gr{\"a}fener}, G. and {Foellmi}, C. and {Schnurr}, O. and {Hamann}, W.-R.},
        title = "{The Wolf-Rayet stars in the Large Magellanic Cloud. A comprehensive analysis of the WN class}",
      journal = {\aap},
         year = 2014,
        month = may,
       volume = {565},
          eid = {A27},
        pages = {A27},
          doi = {10.1051/0004-6361/201322696},
archivePrefix = {arXiv},
       eprint = {1401.5474},
 primaryClass = {astro-ph.SR},
       adsurl = {https://ui.adsabs.harvard.edu/abs/2014A&A...565A..27H}
}

@ARTICLE{Castro2018,
       author = {{Castro}, N. and {Crowther}, P.~A. and {Evans}, C.~J. and {Mackey}, J. and {Castro-Rodriguez}, N. and {Vink}, J.~S. and {Melnick}, J. and {Selman}, F.},
        title = "{Mapping the core of the Tarantula Nebula with VLT-MUSE. I. Spectral and nebular content around R136}",
      journal = {\aap},
         year = 2018,
        month = jun,
       volume = {614},
          eid = {A147},
        pages = {A147},
          doi = {10.1051/0004-6361/201732084},
archivePrefix = {arXiv},
       eprint = {1802.01597},
 primaryClass = {astro-ph.GA},
       adsurl = {https://ui.adsabs.harvard.edu/abs/2018A&A...614A.147C}
}

@ARTICLE{Crowther2023,
       author = {{Crowther}, Paul A. and {Rate}, G. and {Bestenlehner}, Joachim M.},
        title = "{Line luminosities of Galactic and Magellanic Cloud Wolf-Rayet stars}",
      journal = {\mnras},
         year = 2023,
        month = may,
       volume = {521},
       number = {1},
        pages = {585-612},
          doi = {10.1093/mnras/stad418},
archivePrefix = {arXiv},
       eprint = {2301.11297},
 primaryClass = {astro-ph.SR},
       adsurl = {https://ui.adsabs.harvard.edu/abs/2023MNRAS.521..585C}
}

@ARTICLE{Crowther2006,
       author = {{Crowther}, P.~A. and {Hadfield}, L.~J.},
        title = "{Reduced Wolf-Rayet line luminosities at low metallicity}",
      journal = {\aap},
         year = 2006,
        month = apr,
       volume = {449},
       number = {2},
        pages = {711-722},
          doi = {10.1051/0004-6361:20054298},
archivePrefix = {arXiv},
       eprint = {astro-ph/0512183},
 primaryClass = {astro-ph},
       adsurl = {https://ui.adsabs.harvard.edu/abs/2006A&A...449..711C}
}

@ARTICLE{Madau_Maiolino_2026,
       author = {{Madau}, Piero and {Maiolino}, Roberto},
        title = "{Little Red Dots as Obscured Little Blue Dots: A Super-Eddington Unification Model}",
      journal = {arXiv e-prints},
         year = 2026,
        month = feb,
          eid = {arXiv:2602.22386},
        pages = {arXiv:2602.22386},
          doi = {10.48550/arXiv.2602.22386},
archivePrefix = {arXiv},
       eprint = {2602.22386},
 primaryClass = {astro-ph.GA},
       adsurl = {https://ui.adsabs.harvard.edu/abs/2026arXiv260222386M}
}

@ARTICLE{Perna2023,
       author = {{Perna}, M. and {Arribas}, S. and {Marshall}, M. and {D'Eugenio}, F. and {{\"U}bler}, H. and {Bunker}, A. and {Charlot}, S. and {Carniani}, S. and {Jakobsen}, P. and {Maiolino}, R. and {Rodr{\'\i}guez Del Pino}, B. and {Willott}, C.~J. and {B{\"o}ker}, T. and {Circosta}, C. and {Cresci}, G. and {Curti}, M. and {Husemann}, B. and {Kumari}, N. and {Lamperti}, I. and {P{\'e}rez-Gonz{\'a}lez}, P.~G. and {Scholtz}, J.},
        title = "{GA-NIFS: The ultra-dense, interacting environment of a dual AGN at z {\ensuremath{\sim}} 3.3 revealed by JWST/NIRSpec IFS}",
      journal = {\aap},
         year = 2023,
        month = nov,
       volume = {679},
          eid = {A89},
        pages = {A89},
          doi = {10.1051/0004-6361/202346649},
archivePrefix = {arXiv},
       eprint = {2304.06756},
 primaryClass = {astro-ph.GA},
       adsurl = {https://ui.adsabs.harvard.edu/abs/2023A&A...679A..89P}
}

@ARTICLE{Maiolino:2025,
       author = {{Maiolino}, Roberto and {Uebler}, Hannah and {D'Eugenio}, Francesco and {Scholtz}, Jan and {Juodzbalis}, Ignas and {Ji}, Xihan and {Perna}, Michele and {Bromm}, Volker and {Dayal}, Pratika and {Koudmani}, Sophie and et al.},
        title = "{A black hole in a near-pristine galaxy 700 million years after the Big Bang}",
      journal = {arXiv e-prints},
         year = 2025,
        month = may,
          eid = {arXiv:2505.22567},
        pages = {arXiv:2505.22567},
          doi = {10.48550/arXiv.2505.22567},
archivePrefix = {arXiv},
       eprint = {2505.22567},
 primaryClass = {astro-ph.GA},
       adsurl = {https://ui.adsabs.harvard.edu/abs/2025arXiv250522567M}
}

@ARTICLE{Fujimoto:2025,
       author = {{Fujimoto}, Seiji and {Naidu}, Rohan P. and {Chisholm}, John and {Atek}, Hakim and {Endsley}, Ryan and {Kokorev}, Vasily and {Furtak}, Lukas J. and {Pan}, Richard and {Liu}, Boyuan and {Bromm}, Volker and {Venditti}, Alessandra and {Visbal}, Eli and {Sarmento}, Richard and {Weibel}, Andrea and {Oesch}, Pascal A. and {Brammer}, Gabriel and {Schaerer}, Daniel and {Adamo}, Angela and {Berg}, Danielle A. and {Bezanson}, Rachel and {Bouwens}, Rychard and {Chemerynska}, Iryna and {Claeyssens}, Ad{\'e}la{\"\i}de and {Dessauges-Zavadsky}, Miroslava and {Frebel}, Anna and {Korber}, Damien and {Labbe}, Ivo and {Marques-Chaves}, Rui and {Matthee}, Jorryt and {McQuinn}, Kristen B.~W. and {Mu{\~n}oz}, Julian B. and {Natarajan}, Priyamvada and {Saldana-Lopez}, Alberto and {Suess}, Katherine A. and {Volonteri}, Marta and {Zitrin}, Adi},
        title = "{GLIMPSE: An Ultrafaint ≃{}10$^{5}$ M$_{{\ensuremath{\odot}}}$ Pop III Galaxy Candidate and First Constraints on the Pop III UV Luminosity Function at z ≃ 6{\textendash}7}",
      journal = {\apj},
         year = 2025,
        month = aug,
       volume = {989},
       number = {1},
          eid = {46},
        pages = {46},
          doi = {10.3847/1538-4357/ade9a1},
archivePrefix = {arXiv},
       eprint = {2501.11678},
 primaryClass = {astro-ph.GA},
       adsurl = {https://ui.adsabs.harvard.edu/abs/2025ApJ...989...46F}
}

@ARTICLE{Klessen:2023,
       author = {{Klessen}, Ralf S. and {Glover}, Simon C.~O.},
        title = "{The First Stars: Formation, Properties, and Impact}",
      journal = {\araa},
         year = 2023,
        month = aug,
       volume = {61},
        pages = {65-130},
          doi = {10.1146/annurev-astro-071221-053453},
archivePrefix = {arXiv},
       eprint = {2303.12500},
 primaryClass = {astro-ph.CO},
       adsurl = {https://ui.adsabs.harvard.edu/abs/2023ARA&A..61...65K}
}

@ARTICLE{Schaerer:2025,
       author = {{Schaerer}, D. and {Guibert}, J. and {Marques-Chaves}, R. and {Martins}, F.},
        title = "{Observable and ionizing properties of star-forming galaxies with very massive stars and different initial mass functions}",
      journal = {\aap},
         year = 2025,
        month = jan,
       volume = {693},
          eid = {A271},
        pages = {A271},
          doi = {10.1051/0004-6361/202451454},
archivePrefix = {arXiv},
       eprint = {2407.12122},
 primaryClass = {astro-ph.GA},
       adsurl = {https://ui.adsabs.harvard.edu/abs/2025A&A...693A.271S}
}

@ARTICLE{Trussler:2023,
       author = {{Trussler}, James A.~A. and {Conselice}, Christopher J. and {Adams}, Nathan J. and {Maiolino}, Roberto and {Nakajima}, Kimihiko and {Zackrisson}, Erik and {Austin}, Duncan and {Ferreira}, Leonardo and {Harvey}, Tom},
        title = "{On the observability and identification of Population III galaxies with JWST}",
      journal = {\mnras},
         year = 2023,
        month = nov,
       volume = {525},
       number = {4},
        pages = {5328-5352},
          doi = {10.1093/mnras/stad2553},
archivePrefix = {arXiv},
       eprint = {2211.02038},
 primaryClass = {astro-ph.GA},
       adsurl = {https://ui.adsabs.harvard.edu/abs/2023MNRAS.525.5328T}
}

@ARTICLE{Storck:2025,
       author = {{Storck}, Anatole and {Katz}, Harley and {Devriendt}, Julien and {Slyz}, Adrianne and {Cadiou}, Corentin and {Choustikov}, Nicholas and {Rey}, Martin P. and {Saxena}, Aayush and {Agertz}, Oscar and {Kimm}, Taysun},
        title = "{MEGATRON: The environments of Population III stars at Cosmic Dawn and their connection to present day galaxies}",
      journal = {arXiv e-prints},
         year = 2025,
        month = oct,
          eid = {arXiv:2510.06853},
        pages = {arXiv:2510.06853},
          doi = {10.48550/arXiv.2510.06853},
archivePrefix = {arXiv},
       eprint = {2510.06853},
 primaryClass = {astro-ph.GA},
       adsurl = {https://ui.adsabs.harvard.edu/abs/2025arXiv251006853S}
}

@ARTICLE{Wasserman:2026,
       author = {{Wasserman}, Joel and {Zackrisson}, Erik and {Dhandha}, Jiten and {Fialkov}, Anastasia and {Noble}, Leon and {Majumdar}, Suman},
        title = "{Ultraviolet photon production rates of the first stars: Impact on the He II {\ensuremath{\lambda}} 1640 {\r{A}} emission line from primordial star clusters and the 21-cm signal from cosmic dawn}",
      journal = {\mnras},
         year = 2026,
        month = feb,
          doi = {10.1093/mnras/stag386},
archivePrefix = {arXiv},
       eprint = {2507.21764},
 primaryClass = {astro-ph.GA},
       adsurl = {https://ui.adsabs.harvard.edu/abs/2026MNRAS.tmp..379W}
}

@ARTICLE{Maiolino:2024_Halo,
       author = {{Maiolino}, Roberto and {{\"U}bler}, Hannah and {Perna}, Michele and {Scholtz}, Jan and {D'Eugenio}, Francesco and {Witten}, Callum and {Laporte}, Nicolas and {Witstok}, Joris and {Carniani}, Stefano and {Tacchella}, Sandro and et al.},
        title = "{JADES. Possible Population III signatures at z = 10.6 in the halo of GN-z11}",
      journal = {\aap},
         year = 2024,
        month = jul,
       volume = {687},
          eid = {A67},
        pages = {A67},
          doi = {10.1051/0004-6361/202347087},
archivePrefix = {arXiv},
       eprint = {2306.00953},
 primaryClass = {astro-ph.GA},
       adsurl = {https://ui.adsabs.harvard.edu/abs/2024A&A...687A..67M}
}

@ARTICLE{Oesch:2016,
       author = {{Oesch}, P.~A. and {Brammer}, G. and {van Dokkum}, P.~G. and {Illingworth}, G.~D. and {Bouwens}, R.~J. and {Labb{\'e}}, I. and {Franx}, M. and {Momcheva}, I. and {Ashby}, M.~L.~N. and {Fazio}, G.~G. and et al.},
        title = "{A Remarkably Luminous Galaxy at z=11.1 Measured with Hubble Space Telescope Grism Spectroscopy}",
      journal = {\apj},
         year = 2016,
        month = mar,
       volume = {819},
       number = {2},
          eid = {129},
        pages = {129},
          doi = {10.3847/0004-637X/819/2/129},
archivePrefix = {arXiv},
       eprint = {1603.00461},
 primaryClass = {astro-ph.GA},
       adsurl = {https://ui.adsabs.harvard.edu/abs/2016ApJ...819..129O}
}

@ARTICLE{AlvarezMarquez:2025,
       author = {{{\'A}lvarez-M{\'a}rquez}, J. and {Crespo G{\'o}mez}, A. and {Colina}, L. and {Langeroodi}, D. and {Marques-Chaves}, R. and {Prieto-Jim{\'e}nez}, C. and {Bik}, A. and {Alonso-Herrero}, A. and {Boogaard}, L. and {Costantin}, L. and et al.},
        title = "{Insight into the starburst nature of Galaxy GN-z11 with JWST MIRI spectroscopy}",
      journal = {\aap},
         year = 2025,
        month = mar,
       volume = {695},
          eid = {A250},
        pages = {A250},
          doi = {10.1051/0004-6361/202451731},
archivePrefix = {arXiv},
       eprint = {2412.12826},
 primaryClass = {astro-ph.GA},
       adsurl = {https://ui.adsabs.harvard.edu/abs/2025A&A...695A.250A}
}

@ARTICLE{Scholtz:2024,
       author = {{Scholtz}, Jan and {Witten}, Callum and {Laporte}, Nicolas and {{\"U}bler}, Hannah and {Perna}, Michele and {Maiolino}, Roberto and {Arribas}, Santiago and {Baker}, William M. and {Bennett}, Jake S. and {D'Eugenio}, Francesco and et al.},
        title = "{GN-z11: The environment of an active galactic nucleus at z = 10.603. New insights into the most distant Ly{\ensuremath{\alpha}} detection}",
      journal = {\aap},
         year = 2024,
        month = jul,
       volume = {687},
          eid = {A283},
        pages = {A283},
          doi = {10.1051/0004-6361/202347187},
archivePrefix = {arXiv},
       eprint = {2306.09142},
 primaryClass = {astro-ph.GA},
       adsurl = {https://ui.adsabs.harvard.edu/abs/2024A&A...687A.283S}
}

@ARTICLE{Fabian:2026,
       author = {{Fabian}, A.~C. and {Jiang}, J. and {Baker}, W.~M. and {Maiolino}, R. and {Ji}, X. and {Juod{\v{z}}balis}, I. and {Scholtz}, J.},
        title = "{The possible accretion discs of GN-z11 at redshift z = 10.6, MoM-z14 at z = 14.44 and other high redshift objects}",
      journal = {\mnras},
         year = 2026,
        month = feb,
          doi = {10.1093/mnras/stag379},
archivePrefix = {arXiv},
       eprint = {2509.05459},
 primaryClass = {astro-ph.GA},
       adsurl = {https://ui.adsabs.harvard.edu/abs/2026MNRAS.tmp..376F}
}

@ARTICLE{CrespoGomez:2026,
       author = {{Crespo G{\'o}mez}, A. and {Colina}, L. and {P{\'e}rez-Gonz{\'a}lez}, P.~G. and {{\'A}lvarez-M{\'a}rquez}, J. and {Garc{\'\i}a-Mar{\'\i}n}, M. and {Alonso-Herrero}, A. and {Annunziatella}, M. and {Bik}, A. and {Bosman}, S. and {Bunker}, A.~J. and et al.},
        title = "{MIRI spectrophotometry of GN-z11: Detection and nature of an optical red continuum component}",
      journal = {\aap},
         year = 2026,
        month = jan,
       volume = {706},
          eid = {A46},
        pages = {A46},
          doi = {10.1051/0004-6361/202556814},
archivePrefix = {arXiv},
       eprint = {2512.02997},
 primaryClass = {astro-ph.GA},
       adsurl = {https://ui.adsabs.harvard.edu/abs/2026A&A...706A..46C}
}

@ARTICLE{Venditti:2023,
       author = {{Venditti}, Alessandra and {Graziani}, Luca and {Schneider}, Raffaella and {Pentericci}, Laura and {Di Cesare}, Claudia and {Maio}, Umberto and {Omukai}, Kazuyuki},
        title = "{A needle in a haystack? Catching Population III stars in the epoch of reionization: I. Population III star-forming environments}",
      journal = {\mnras},
         year = 2023,
        month = jul,
       volume = {522},
       number = {3},
        pages = {3809-3830},
          doi = {10.1093/mnras/stad1201},
archivePrefix = {arXiv},
       eprint = {2301.10259},
 primaryClass = {astro-ph.GA},
       adsurl = {https://ui.adsabs.harvard.edu/abs/2023MNRAS.522.3809V}
}

@ARTICLE{Venditti:2024,
       author = {{Venditti}, Alessandra and {Bromm}, Volker and {Finkelstein}, Steven L. and {Calabr{\`o}}, Antonello and {Napolitano}, Lorenzo and {Graziani}, Luca and {Schneider}, Raffaella},
        title = "{A Hide-and-seek Game: Looking for Population III Stars during the Epoch of Reionization through the He II {\ensuremath{\lambda}}1640 Line}",
      journal = {\apjl},
         year = 2024,
        month = sep,
       volume = {973},
       number = {1},
          eid = {L12},
        pages = {L12},
          doi = {10.3847/2041-8213/ad7387},
archivePrefix = {arXiv},
       eprint = {2405.10940},
 primaryClass = {astro-ph.GA},
       adsurl = {https://ui.adsabs.harvard.edu/abs/2024ApJ...973L..12V}
}

@ARTICLE{Venditti:2025,
       author = {{Venditti}, Alessandra and {Mu{\~n}oz}, Julian B. and {Bromm}, Volker and {Fujimoto}, Seiji and {Finkelstein}, Steven L. and {Chisholm}, John},
        title = "{Bursty or Heavy? The Surprise of Bright Population III Systems in the Reionization Era}",
      journal = {\apj},
         year = 2025,
        month = nov,
       volume = {994},
       number = {1},
          eid = {32},
        pages = {32},
          doi = {10.3847/1538-4357/ae0610},
archivePrefix = {arXiv},
       eprint = {2505.20263},
 primaryClass = {astro-ph.GA},
       adsurl = {https://ui.adsabs.harvard.edu/abs/2025ApJ...994...32V}
}

@ARTICLE{Liu:2020,
       author = {{Liu}, Boyuan and {Bromm}, Volker},
        title = "{When did Population III star formation end?}",
      journal = {\mnras},
         year = 2020,
        month = sep,
       volume = {497},
       number = {3},
        pages = {2839-2854},
          doi = {10.1093/mnras/staa2143},
archivePrefix = {arXiv},
       eprint = {2006.15260},
 primaryClass = {astro-ph.GA},
       adsurl = {https://ui.adsabs.harvard.edu/abs/2020MNRAS.497.2839L}
}

@ARTICLE{Shajib:2025,
       author = {{Shajib}, Anowar J. and {Treu}, Tommaso and {Melo}, Alejandra and {Roberts-Borsani}, Guido and {Knabel}, Shawn and {Cappellari}, Michele and {Frieman}, Joshua A.},
        title = "{An accurate measurement of the spectral resolution of the JWST Near Infrared Spectrograph}",
      journal = {\aap},
         year = 2025,
        month = oct,
       volume = {702},
          eid = {L12},
        pages = {L12},
          doi = {10.1051/0004-6361/202556281},
archivePrefix = {arXiv},
       eprint = {2507.03746},
 primaryClass = {astro-ph.IM},
       adsurl = {https://ui.adsabs.harvard.edu/abs/2025A&A...702L..12S}
}

@ARTICLE{Rusta:2025,
       author = {{Rusta}, Elka and {Salvadori}, Stefania and {Gelli}, Viola and {Schaerer}, Daniel and {Marconi}, Alessandro and {Koutsouridou}, Ioanna and {Carniani}, Stefano},
        title = "{Metal-polluted Population III Galaxies and How to Find Them}",
      journal = {\apjl},
         year = 2025,
        month = aug,
       volume = {989},
       number = {2},
          eid = {L32},
        pages = {L32},
          doi = {10.3847/2041-8213/adf4e3},
archivePrefix = {arXiv},
       eprint = {2506.17400},
 primaryClass = {astro-ph.GA},
       adsurl = {https://ui.adsabs.harvard.edu/abs/2025ApJ...989L..32R}
}

@ARTICLE{Schneider:2024,
       author = {{Schneider}, Raffaella and {Maiolino}, Roberto},
        title = "{The formation and cosmic evolution of dust in the early Universe: I. Dust sources}",
      journal = {\aapr},
         year = 2024,
        month = apr,
       volume = {32},
       number = {1},
          eid = {2},
        pages = {2},
          doi = {10.1007/s00159-024-00151-2},
archivePrefix = {arXiv},
       eprint = {2310.00053},
 primaryClass = {astro-ph.GA},
       adsurl = {https://ui.adsabs.harvard.edu/abs/2024A&ARv..32....2S}
}

@ARTICLE{Rakshit:2020,
       author = {{Rakshit}, Suvendu and {Stalin}, C.~S. and {Kotilainen}, Jari},
        title = "{Spectral Properties of Quasars from Sloan Digital Sky Survey Data Release 14: The Catalog}",
      journal = {\apjs},
         year = 2020,
        month = jul,
       volume = {249},
       number = {1},
          eid = {17},
        pages = {17},
          doi = {10.3847/1538-4365/ab99c5},
archivePrefix = {arXiv},
       eprint = {1910.10395},
 primaryClass = {astro-ph.GA},
       adsurl = {https://ui.adsabs.harvard.edu/abs/2020ApJS..249...17R}
}

@ARTICLE{HamelBravo:2025,
       author = {{Hamel-Bravo}, M.~J. and {Fisher}, D.~B. and {Berg}, D.~A. and {Cameron}, A.~J. and {Chisholm}, J. and {Kacprzak}, G.~G. and {Mazzilli Ciraulo}, B. and {Katz}, H.},
        title = "{Declining metallicity and extended He II in the outflow of an Epoch of Reionisation analogue galaxy}",
      journal = {\aap},
         year = 2025,
        month = nov,
       volume = {704},
          eid = {L4},
        pages = {L4},
          doi = {10.1051/0004-6361/202557538},
archivePrefix = {arXiv},
       eprint = {2510.05332},
 primaryClass = {astro-ph.GA},
       adsurl = {https://ui.adsabs.harvard.edu/abs/2025A&A...704L...4H}
}

@ARTICLE{2010A&A...523A..64R,
       author = {{Raiter}, A. and {Schaerer}, D. and {Fosbury}, R.~A.~E.},
        title = "{Predicted UV properties of very metal-poor starburst galaxies}",
      journal = {\aap},
         year = 2010,
        month = nov,
       volume = {523},
          eid = {A64},
        pages = {A64},
          doi = {10.1051/0004-6361/201015236},
archivePrefix = {arXiv},
       eprint = {1008.2114},
 primaryClass = {astro-ph.CO},
       adsurl = {https://ui.adsabs.harvard.edu/abs/2010A&A...523A..64R}
}

@ARTICLE{2013Natur.502..524F,
       author = {{Finkelstein}, S.~L. and {Papovich}, C. and {Dickinson}, M. and {Song}, M. and {Tilvi}, V. and {Koekemoer}, A.~M. and {Finkelstein}, K.~D. and {Mobasher}, B. and {Ferguson}, H.~C. and {Giavalisco}, M. and et al.},
        title = "{A galaxy rapidly forming stars 700 million years after the Big Bang at redshift 7.51}",
      journal = {\nat},
         year = 2013,
        month = oct,
       volume = {502},
       number = {7472},
        pages = {524-527},
          doi = {10.1038/nature12657},
archivePrefix = {arXiv},
       eprint = {1310.6031},
 primaryClass = {astro-ph.CO},
       adsurl = {https://ui.adsabs.harvard.edu/abs/2013Natur.502..524F}
}

@ARTICLE{2015ApJ...804L..30O,
       author = {{Oesch}, P.~A. and {van Dokkum}, P.~G. and {Illingworth}, G.~D. and {Bouwens}, R.~J. and {Momcheva}, I. and {Holden}, B. and {Roberts-Borsani}, G.~W. and {Smit}, R. and {Franx}, M. and {Labb{\'e}}, I. and et al.},
        title = "{A Spectroscopic Redshift Measurement for a Luminous Lyman Break Galaxy at z = 7.730 Using Keck/MOSFIRE}",
      journal = {\apjl},
         year = 2015,
        month = may,
       volume = {804},
       number = {2},
          eid = {L30},
        pages = {L30},
          doi = {10.1088/2041-8205/804/2/L30},
archivePrefix = {arXiv},
       eprint = {1502.05399},
 primaryClass = {astro-ph.GA},
       adsurl = {https://ui.adsabs.harvard.edu/abs/2015ApJ...804L..30O}
}

@ARTICLE{2015ApJ...810L..12Z,
       author = {{Zitrin}, Adi and {Labb{\'e}}, Ivo and {Belli}, Sirio and {Bouwens}, Rychard and {Ellis}, Richard S. and {Roberts-Borsani}, Guido and {Stark}, Daniel P. and {Oesch}, Pascal A. and {Smit}, Renske},
        title = "{Lyman{\ensuremath{\alpha}} Emission from a Luminous z = 8.68 Galaxy: Implications for Galaxies as Tracers of Cosmic Reionization}",
      journal = {\apjl},
         year = 2015,
        month = sep,
       volume = {810},
       number = {1},
          eid = {L12},
        pages = {L12},
          doi = {10.1088/2041-8205/810/1/L12},
archivePrefix = {arXiv},
       eprint = {1507.02679},
 primaryClass = {astro-ph.GA},
       adsurl = {https://ui.adsabs.harvard.edu/abs/2015ApJ...810L..12Z}
}

@ARTICLE{2016ApJ...823..143R,
       author = {{Roberts-Borsani}, G.~W. and {Bouwens}, R.~J. and {Oesch}, P.~A. and {Labbe}, I. and {Smit}, R. and {Illingworth}, G.~D. and {van Dokkum}, P. and {Holden}, B. and {Gonzalez}, V. and {Stefanon}, M. and et al.},
        title = "{z {\ensuremath{\gtrsim}} 7 Galaxies with Red Spitzer/IRAC [3.6]-[4.5] Colors in the Full CANDELS Data Set: The Brightest-Known Galaxies at z \raisebox{-0.5ex}\textasciitilde 7-9 and a Probable Spectroscopic Confirmation at z = 7.48}",
      journal = {\apj},
         year = 2016,
        month = jun,
       volume = {823},
       number = {2},
          eid = {143},
        pages = {143},
          doi = {10.3847/0004-637X/823/2/143},
archivePrefix = {arXiv},
       eprint = {1506.00854},
 primaryClass = {astro-ph.GA},
       adsurl = {https://ui.adsabs.harvard.edu/abs/2016ApJ...823..143R}
}

@ARTICLE{2018ApJ...856....2M,
       author = {{Mason}, Charlotte A. and {Treu}, Tommaso and {Dijkstra}, Mark and {Mesinger}, Andrei and {Trenti}, Michele and {Pentericci}, Laura and {de Barros}, Stephane and {Vanzella}, Eros},
        title = "{The Universe Is Reionizing at z {\ensuremath{\sim}} 7: Bayesian Inference of the IGM Neutral Fraction Using Ly{\ensuremath{\alpha}} Emission from Galaxies}",
      journal = {\apj},
         year = 2018,
        month = mar,
       volume = {856},
       number = {1},
          eid = {2},
        pages = {2},
          doi = {10.3847/1538-4357/aab0a7},
archivePrefix = {arXiv},
       eprint = {1709.05356},
 primaryClass = {astro-ph.CO},
       adsurl = {https://ui.adsabs.harvard.edu/abs/2018ApJ...856....2M}
}

@ARTICLE{2020MNRAS.499.1395M,
       author = {{Mason}, Charlotte A. and {Gronke}, Max},
        title = "{Measuring the properties of reionized bubbles with resolved Ly{\ensuremath{\alpha}} spectra}",
      journal = {\mnras},
         year = 2020,
        month = nov,
       volume = {499},
       number = {1},
        pages = {1395-1405},
          doi = {10.1093/mnras/staa2910},
archivePrefix = {arXiv},
       eprint = {2004.13065},
 primaryClass = {astro-ph.GA},
       adsurl = {https://ui.adsabs.harvard.edu/abs/2020MNRAS.499.1395M}
}

@ARTICLE{2022ApJ...930..104L,
       author = {{Larson}, Rebecca L. and {Finkelstein}, Steven L. and {Hutchison}, Taylor A. and {Papovich}, Casey and {Bagley}, Micaela and {Dickinson}, Mark and {Rojas-Ruiz}, Sof{\'\i}a and {Ferguson}, Harry C. and {Jung}, Intae and {Giavalisco}, Mauro and et al.},
        title = "{Searching for Islands of Reionization: A Potential Ionized Bubble Powered by a Spectroscopic Overdensity at z = 8.7}",
      journal = {\apj},
         year = 2022,
        month = may,
       volume = {930},
       number = {2},
          eid = {104},
        pages = {104},
          doi = {10.3847/1538-4357/ac5dbd},
archivePrefix = {arXiv},
       eprint = {2203.08461},
 primaryClass = {astro-ph.GA},
       adsurl = {https://ui.adsabs.harvard.edu/abs/2022ApJ...930..104L}
}

@ARTICLE{2025MNRAS.536...27W,
       author = {{Witstok}, Joris and {Maiolino}, Roberto and {Smit}, Renske and {Jones}, Gareth C. and {Bunker}, Andrew J. and {Helton}, Jakob M. and {Johnson}, Benjamin D. and {Tacchella}, Sandro and {Saxena}, Aayush and {Arribas}, Santiago and et al.},
        title = "{JADES: primaeval Lyman {\ensuremath{\alpha}} emitting galaxies reveal early sites of reionization out to redshift z \raisebox{-0.5ex}\textasciitilde 9}",
      journal = {\mnras},
         year = 2025,
        month = jan,
       volume = {536},
       number = {1},
        pages = {27-50},
          doi = {10.1093/mnras/stae2535},
archivePrefix = {arXiv},
       eprint = {2404.05724},
 primaryClass = {astro-ph.GA},
       adsurl = {https://ui.adsabs.harvard.edu/abs/2025MNRAS.536...27W}
}

@ARTICLE{2024MNRAS.535.1796B,
       author = {{Boyett}, Kit and {Bunker}, Andrew J. and {Curtis-Lake}, Emma and {Chevallard}, Jacopo and {Cameron}, Alex J. and {Jones}, Gareth C. and {Saxena}, Aayush and {Charlot}, St{\'e}phane and {Curti}, Mirko and {Wallace}, Imaan E.~B. and et al.},
        title = "{Extreme emission line galaxies detected in JADES JWST/NIRSpec - I. Inferred galaxy properties}",
      journal = {\mnras},
         year = 2024,
        month = dec,
       volume = {535},
       number = {2},
        pages = {1796-1828},
          doi = {10.1093/mnras/stae2430},
archivePrefix = {arXiv},
       eprint = {2401.16934},
 primaryClass = {astro-ph.GA},
       adsurl = {https://ui.adsabs.harvard.edu/abs/2024MNRAS.535.1796B}
}

@ARTICLE{2023MNRAS.526.1657T,
       author = {{Tang}, Mengtao and {Stark}, Daniel P. and {Chen}, Zuyi and {Mason}, Charlotte and {Topping}, Michael and {Endsley}, Ryan and {Senchyna}, Peter and {Plat}, Ad{\`e}le and {Lu}, Ting-Yi and {Whitler}, Lily and et al.},
        title = "{JWST/NIRSpec spectroscopy of z = 7-9 star-forming galaxies with CEERS: new insight into bright Ly{\ensuremath{\alpha}} emitters in ionized bubbles}",
      journal = {\mnras},
         year = 2023,
        month = dec,
       volume = {526},
       number = {2},
        pages = {1657-1686},
          doi = {10.1093/mnras/stad2763},
archivePrefix = {arXiv},
       eprint = {2301.07072},
 primaryClass = {astro-ph.GA},
       adsurl = {https://ui.adsabs.harvard.edu/abs/2023MNRAS.526.1657T}
}

@ARTICLE{2024ApJ...970...50C,
       author = {{Cooper}, Olivia R. and {Casey}, Caitlin M. and {Akins}, Hollis B. and {Magee}, Jake and {Melendez}, Alfonso and {Fong}, Mia and {Urbano Stawinski}, Stephanie M. and {Kartaltepe}, Jeyhan S. and {Finkelstein}, Steven L. and {Larson}, Rebecca L. and et al.},
        title = "{The Web Epoch of Reionization Ly{\ensuremath{\alpha}} Survey (WERLS). I. MOSFIRE Spectroscopy of z {\ensuremath{\sim}} 7{\textendash}8 Ly{\ensuremath{\alpha}} Emitters}",
      journal = {\apj},
         year = 2024,
        month = jul,
       volume = {970},
       number = {1},
          eid = {50},
        pages = {50},
          doi = {10.3847/1538-4357/ad4c6c},
archivePrefix = {arXiv},
       eprint = {2309.06656},
 primaryClass = {astro-ph.GA},
       adsurl = {https://ui.adsabs.harvard.edu/abs/2024ApJ...970...50C}
}

@ARTICLE{2025arXiv251206072L,
       author = {{Leonova}, Ecaterina and {Naidu}, Rohan P. and {Oesch}, Pascal A. and {Brammer}, Gabriel and {Matthee}, Jorryt and {Meyer}, Romain A. and {Schaerer}, Daniel and {Xiao}, Mengyuan},
        title = "{Lyman-$α$ Visibility During the Epoch of Reionization: Combining JWST FRESCO Grism Data with Keck Archival Spectroscopy}",
      journal = {arXiv e-prints},
         year = 2025,
        month = dec,
          eid = {arXiv:2512.06072},
        pages = {arXiv:2512.06072},
archivePrefix = {arXiv},
       eprint = {2512.06072},
 primaryClass = {astro-ph.GA},
       adsurl = {https://ui.adsabs.harvard.edu/abs/2025arXiv251206072L}
}

@ARTICLE{2024A&A...687A.283S,
       author = {{Scholtz}, Jan and {Witten}, Callum and {Laporte}, Nicolas and {{\"U}bler}, Hannah and {Perna}, Michele and {Maiolino}, Roberto and {Arribas}, Santiago and {Baker}, William M. and {Bennett}, Jake S. and {D'Eugenio}, Francesco and et al.},
        title = "{GN-z11: The environment of an active galactic nucleus at z = 10.603. New insights into the most distant Ly{\ensuremath{\alpha}} detection}",
      journal = {\aap},
         year = 2024,
        month = jul,
       volume = {687},
          eid = {A283},
        pages = {A283},
          doi = {10.1051/0004-6361/202347187},
archivePrefix = {arXiv},
       eprint = {2306.09142},
 primaryClass = {astro-ph.GA},
       adsurl = {https://ui.adsabs.harvard.edu/abs/2024A&A...687A.283S}
}

@ARTICLE{2024A&A...683A.238J,
       author = {{Jones}, Gareth C. and {Bunker}, Andrew J. and {Saxena}, Aayush and {Witstok}, Joris and {Stark}, Daniel P. and {Arribas}, Santiago and {Baker}, William M. and {Bhatawdekar}, Rachana and {Bowler}, Rebecca and {Boyett}, Kristan and et al.},
        title = "{JADES: The emergence and evolution of Ly{\ensuremath{\alpha}} emission and constraints on the intergalactic medium neutral fraction}",
      journal = {\aap},
         year = 2024,
        month = mar,
       volume = {683},
          eid = {A238},
        pages = {A238},
          doi = {10.1051/0004-6361/202347099},
archivePrefix = {arXiv},
       eprint = {2306.02471},
 primaryClass = {astro-ph.GA},
       adsurl = {https://ui.adsabs.harvard.edu/abs/2024A&A...683A.238J}
}

@ARTICLE{2024ApJ...975..208T,
       author = {{Tang}, Mengtao and {Stark}, Daniel P. and {Topping}, Michael W. and {Mason}, Charlotte and {Ellis}, Richard S.},
        title = "{JWST/NIRSpec Observations of Lyman {\ensuremath{\alpha}} Emission in Star-forming Galaxies at 6.5 {\ensuremath{\lesssim}} z {\ensuremath{\lesssim}} 13}",
      journal = {\apj},
         year = 2024,
        month = nov,
       volume = {975},
       number = {2},
          eid = {208},
        pages = {208},
          doi = {10.3847/1538-4357/ad7eb7},
archivePrefix = {arXiv},
       eprint = {2408.01507},
 primaryClass = {astro-ph.GA},
       adsurl = {https://ui.adsabs.harvard.edu/abs/2024ApJ...975..208T}
}

@ARTICLE{2025Natur.639..897W,
       author = {{Witstok}, Joris and {Jakobsen}, Peter and {Maiolino}, Roberto and {Helton}, Jakob M. and {Johnson}, Benjamin D. and {Robertson}, Brant E. and {Tacchella}, Sandro and {Cameron}, Alex J. and {Smit}, Renske and {Bunker}, Andrew J. and et al.},
        title = "{Witnessing the onset of reionization through Lyman-{\ensuremath{\alpha}} emission at redshift 13}",
      journal = {\nat},
         year = 2025,
        month = mar,
       volume = {639},
       number = {8056},
        pages = {897-901},
          doi = {10.1038/s41586-025-08779-5},
archivePrefix = {arXiv},
       eprint = {2408.16608},
 primaryClass = {astro-ph.GA},
       adsurl = {https://ui.adsabs.harvard.edu/abs/2025Natur.639..897W}
}

@ARTICLE{2025ApJS..278...33K,
       author = {{Kageura}, Yuta and {Ouchi}, Masami and {Nakane}, Minami and {Umeda}, Hiroya and {Harikane}, Yuichi and {Yoshiura}, Shintaro and {Nakajima}, Kimihiko and {Yajima}, Hidenobu and {Thai}, Tran Thi},
        title = "{Census of Ly{\ensuremath{\alpha}} Emission from {\ensuremath{\sim}}600 Galaxies at z = 5{\textendash}14: Evolution of the Ly{\ensuremath{\alpha}} Luminosity Function and a Late Sharp Cosmic Reionization}",
      journal = {\apjs},
         year = 2025,
        month = jun,
       volume = {278},
       number = {2},
          eid = {33},
        pages = {33},
          doi = {10.3847/1538-4365/adc690},
archivePrefix = {arXiv},
       eprint = {2501.05834},
 primaryClass = {astro-ph.GA},
       adsurl = {https://ui.adsabs.harvard.edu/abs/2025ApJS..278...33K}
}

@ARTICLE{2012ApJ...744...83O,
       author = {{Ono}, Yoshiaki and {Ouchi}, Masami and {Mobasher}, Bahram and {Dickinson}, Mark and {Penner}, Kyle and {Shimasaku}, Kazuhiro and {Weiner}, Benjamin J. and {Kartaltepe}, Jeyhan S. and {Nakajima}, Kimihiko and {Nayyeri}, Hooshang and {Stern}, Daniel and {Kashikawa}, Nobunari and {Spinrad}, Hyron},
        title = "{Spectroscopic Confirmation of Three z-dropout Galaxies at z = 6.844-7.213: Demographics of Ly{\ensuremath{\alpha}} Emission in z \raisebox{-0.5ex}\textasciitilde 7 Galaxies}",
      journal = {\apj},
         year = 2012,
        month = jan,
       volume = {744},
       number = {2},
          eid = {83},
        pages = {83},
          doi = {10.1088/0004-637X/744/2/83},
archivePrefix = {arXiv},
       eprint = {1107.3159},
 primaryClass = {astro-ph.CO},
       adsurl = {https://ui.adsabs.harvard.edu/abs/2012ApJ...744...83O}
}

@ARTICLE{Lecroq2025,
       author = {{Lecroq}, Marie and {Charlot}, St{\'e}phane and {Bressan}, Alessandro and {Bruzual}, Gustavo and {Costa}, Guglielmo and {Iorio}, Giuliano and {Mapelli}, Michela and {Santoliquido}, Filippo and {Shepherd}, Kendall and {Spera}, Mario},
        title = "{A new prescription for the spectral properties of population III stellar populations}",
      journal = {\aap},
         year = 2025,
        month = mar,
       volume = {695},
          eid = {A17},
        pages = {A17},
          doi = {10.1051/0004-6361/202452463},
archivePrefix = {arXiv},
       eprint = {2502.14028},
 primaryClass = {astro-ph.GA},
       adsurl = {https://ui.adsabs.harvard.edu/abs/2025A&A...695A..17L}
}

@ARTICLE{Boker2023,
       author = {{B{\"o}ker}, T. and {Beck}, T.~L. and {Birkmann}, S.~M. and {Giardino}, G. and {Keyes}, C. and {Kumari}, N. and {Muzerolle}, J. and {Rawle}, T. and {Zeidler}, P. and {Abul-Huda}, Y. and {Alves de Oliveira}, C. and {Arribas}, S. and {Bechtold}, K. and {Bhatawdekar}, R. and {Bonaventura}, N. and {Bunker}, A.~J. and {Cameron}, A.~J. and {Carniani}, S. and {Charlot}, S. and {Curti}, M. and {Espinoza}, N. and {Ferruit}, P. and {Franx}, M. and {Jakobsen}, P. and {Karakla}, D. and {L{\'o}pez-Caniego}, M. and {L{\"u}tzgendorf}, N. and {Maiolino}, R. and {Manjavacas}, E. and {Marston}, A.~P. and {Moseley}, S.~H. and {Ogle}, P. and {Perna}, M. and {Pe{\~n}a-Guerrero}, M. and {Pirzkal}, N. and {Plesha}, R. and {Proffitt}, C.~R. and {Rauscher}, B.~J. and {Rix}, H. -W. and {Rodr{\'\i}guez del Pino}, B. and {Rustamkulov}, Z. and {Sabbi}, E. and {Sing}, D.~K. and {Sirianni}, M. and {te Plate}, M. and {{\'U}beda}, L. and {Wahlgren}, G.~M. and {Wislowski}, E. and {Wu}, R. and {Willott}, Chris J.},
        title = "{In-orbit Performance of the Near-infrared Spectrograph NIRSpec on the James Webb Space Telescope}",
      journal = {\pasp},
         year = 2023,
        month = mar,
       volume = {135},
       number = {1045},
          eid = {038001},
        pages = {038001},
          doi = {10.1088/1538-3873/acb846},
archivePrefix = {arXiv},
       eprint = {2301.13766},
 primaryClass = {astro-ph.IM},
       adsurl = {https://ui.adsabs.harvard.edu/abs/2023PASP..135c8001B}
}

@ARTICLE{Jakobsen2022,
       author = {{Jakobsen}, P. and {Ferruit}, P. and {Alves de Oliveira}, C. and {Arribas}, S. and {Bagnasco}, G. and {Barho}, R. and {Beck}, T.~L. and {Birkmann}, S. and {B{\"o}ker}, T. and {Bunker}, A.~J. and {Charlot}, S. and {de Jong}, P. and {de Marchi}, G. and {Ehrenwinkler}, R. and {Falcolini}, M. and {Fels}, R. and {Franx}, M. and {Franz}, D. and {Funke}, M. and {Giardino}, G. and {Gnata}, X. and {Holota}, W. and {Honnen}, K. and {Jensen}, P.~L. and {Jentsch}, M. and {Johnson}, T. and {Jollet}, D. and {Karl}, H. and {Kling}, G. and {K{\"o}hler}, J. and {Kolm}, M. -G. and {Kumari}, N. and {Lander}, M.~E. and {Lemke}, R. and {L{\'o}pez-Caniego}, M. and {L{\"u}tzgendorf}, N. and {Maiolino}, R. and {Manjavacas}, E. and {Marston}, A. and {Maschmann}, M. and {Maurer}, R. and {Messerschmidt}, B. and {Moseley}, S.~H. and {Mosner}, P. and {Mott}, D.~B. and {Muzerolle}, J. and {Pirzkal}, N. and {Pittet}, J. -F. and {Plitzke}, A. and {Posselt}, W. and {Rapp}, B. and {Rauscher}, B.~J. and {Rawle}, T. and {Rix}, H. -W. and {R{\"o}del}, A. and {Rumler}, P. and {Sabbi}, E. and {Salvignol}, J. -C. and {Schmid}, T. and {Sirianni}, M. and {Smith}, C. and {Strada}, P. and {te Plate}, M. and {Valenti}, J. and {Wettemann}, T. and {Wiehe}, T. and {Wiesmayer}, M. and {Willott}, C.~J. and {Wright}, R. and {Zeidler}, P. and {Zincke}, C.},
        title = "{The Near-Infrared Spectrograph (NIRSpec) on the James Webb Space Telescope. I. Overview of the instrument and its capabilities}",
      journal = {\aap},
         year = 2022,
        month = may,
       volume = {661},
          eid = {A80},
        pages = {A80},
          doi = {10.1051/0004-6361/202142663},
archivePrefix = {arXiv},
       eprint = {2202.03305},
 primaryClass = {astro-ph.IM},
       adsurl = {https://ui.adsabs.harvard.edu/abs/2022A&A...661A..80J}
}

@INCOLLECTION{Escriva2024,
       author = {{Escriv{\`a}}, Albert and {K{\"u}hnel}, Florian and {Tada}, Yuichiro},
        title = "{Primordial black holes}",
    booktitle = {Black Holes in the Era of Gravitational-Wave Astronomy},
         year = 2024,
       editor = {{Arca Sedda}, Manuel and {Bortolas}, Elisa and {Spera}, Mario},
       publisher = {Elsevier},
        pages = {261-377},
          doi = {10.1016/B978-0-32-395636-9.00012-8},
       adsurl = {https://ui.adsabs.harvard.edu/abs/2024bheg.book..261E}
}

@ARTICLE{Maiolino2024,
       author = {{Maiolino}, Roberto and {Scholtz}, Jan and {Witstok}, Joris and {Carniani}, Stefano and {D'Eugenio}, Francesco and {de Graaff}, Anna and {{\"U}bler}, Hannah and {Tacchella}, Sandro and {Curtis-Lake}, Emma and {Arribas}, Santiago and {Bunker}, Andrew and {Charlot}, St{\'e}phane and {Chevallard}, Jacopo and {Curti}, Mirko and {Looser}, Tobias J. and {Maseda}, Michael V. and {Rawle}, Timothy D. and {Rodr{\'\i}guez del Pino}, Bruno and {Willott}, Chris J. and {Egami}, Eiichi and {Eisenstein}, Daniel J. and {Hainline}, Kevin N. and {Robertson}, Brant and {Williams}, Christina C. and {Willmer}, Christopher N.~A. and {Baker}, William M. and {Boyett}, Kristan and {DeCoursey}, Christa and {Fabian}, Andrew C. and {Helton}, Jakob M. and {Ji}, Zhiyuan and {Jones}, Gareth C. and {Kumari}, Nimisha and {Laporte}, Nicolas and {Nelson}, Erica J. and {Perna}, Michele and {Sandles}, Lester and {Shivaei}, Irene and {Sun}, Fengwu},
        title = "{A small and vigorous black hole in the early Universe}",
      journal = {\nat},
         year = 2024,
        month = mar,
       volume = {627},
       number = {8002},
        pages = {59-63},
          doi = {10.1038/s41586-024-07052-5},
archivePrefix = {arXiv},
       eprint = {2305.12492},
 primaryClass = {astro-ph.GA},
       adsurl = {https://ui.adsabs.harvard.edu/abs/2024Natur.627...59M}
}

@ARTICLE{Tacchella2023,
       author = {{Tacchella}, Sandro and {Eisenstein}, Daniel J. and {Hainline}, Kevin and {Johnson}, Benjamin D. and {Baker}, William M. and {Helton}, Jakob M. and {Robertson}, Brant and {Suess}, Katherine A. and {Chen}, Zuyi and {Nelson}, Erica and {Pusk{\'a}s}, D{\'a}vid and {Sun}, Fengwu and {Alberts}, Stacey and {Egami}, Eiichi and {Hausen}, Ryan and {Rieke}, George and {Rieke}, Marcia and {Shivaei}, Irene and {Williams}, Christina C. and {Willmer}, Christopher N.~A. and {Bunker}, Andrew and {Cameron}, Alex J. and {Carniani}, Stefano and {Charlot}, Stephane and {Curti}, Mirko and {Curtis-Lake}, Emma and {Looser}, Tobias J. and {Maiolino}, Roberto and {Maseda}, Michael V. and {Rawle}, Tim and {Rix}, Hans-Walter and {Smit}, Renske and {{\"U}bler}, Hannah and {Willott}, Chris and {Witstok}, Joris and {Baum}, Stefi and {Bhatawdekar}, Rachana and {Boyett}, Kristan and {Danhaive}, A. Lola and {de Graaff}, Anna and {Endsley}, Ryan and {Ji}, Zhiyuan and {Lyu}, Jianwei and {Sandles}, Lester and {Saxena}, Aayush and {Scholtz}, Jan and {Topping}, Michael W. and {Whitler}, Lily},
        title = "{JADES Imaging of GN-z11: Revealing the Morphology and Environment of a Luminous Galaxy 430 Myr after the Big Bang}",
      journal = {\apj},
         year = 2023,
        month = jul,
       volume = {952},
       number = {1},
          eid = {74},
        pages = {74},
          doi = {10.3847/1538-4357/acdbc6},
archivePrefix = {arXiv},
       eprint = {2302.07234},
 primaryClass = {astro-ph.GA},
       adsurl = {https://ui.adsabs.harvard.edu/abs/2023ApJ...952...74T}
}

@ARTICLE{Bunker2023,
       author = {{Bunker}, Andrew J. and {Saxena}, Aayush and {Cameron}, Alex J. and {Willott}, Chris J. and {Curtis-Lake}, Emma and {Jakobsen}, Peter and {Carniani}, Stefano and {Smit}, Renske and {Maiolino}, Roberto and {Witstok}, Joris and {Curti}, Mirko and {D'Eugenio}, Francesco and {Jones}, Gareth C. and {Ferruit}, Pierre and {Arribas}, Santiago and {Charlot}, Stephane and {Chevallard}, Jacopo and {Giardino}, Giovanna and {de Graaff}, Anna and {Looser}, Tobias J. and {L{\"u}tzgendorf}, Nora and {Maseda}, Michael V. and {Rawle}, Tim and {Rix}, Hans-Walter and {Del Pino}, Bruno Rodr{\'\i}guez and {Alberts}, Stacey and {Egami}, Eiichi and {Eisenstein}, Daniel J. and {Endsley}, Ryan and {Hainline}, Kevin and {Hausen}, Ryan and {Johnson}, Benjamin D. and {Rieke}, George and {Rieke}, Marcia and {Robertson}, Brant E. and {Shivaei}, Irene and {Stark}, Daniel P. and {Sun}, Fengwu and {Tacchella}, Sandro and {Tang}, Mengtao and {Williams}, Christina C. and {Willmer}, Christopher N.~A. and {Baker}, William M. and {Baum}, Stefi and {Bhatawdekar}, Rachana and {Bowler}, Rebecca and {Boyett}, Kristan and {Chen}, Zuyi and {Circosta}, Chiara and {Helton}, Jakob M. and {Ji}, Zhiyuan and {Kumari}, Nimisha and {Lyu}, Jianwei and {Nelson}, Erica and {Parlanti}, Eleonora and {Perna}, Michele and {Sandles}, Lester and {Scholtz}, Jan and {Suess}, Katherine A. and {Topping}, Michael W. and {{\"U}bler}, Hannah and {Wallace}, Imaan E.~B. and {Whitler}, Lily},
        title = "{JADES NIRSpec Spectroscopy of GN-z11: Lyman-{\ensuremath{\alpha}} emission and possible enhanced nitrogen abundance in a z = 10.60 luminous galaxy}",
      journal = {\aap},
         year = 2023,
        month = sep,
       volume = {677},
          eid = {A88},
        pages = {A88},
          doi = {10.1051/0004-6361/202346159},
archivePrefix = {arXiv},
       eprint = {2302.07256},
 primaryClass = {astro-ph.GA},
       adsurl = {https://ui.adsabs.harvard.edu/abs/2023A&A...677A..88B}
}

@ARTICLE{Ubler2023,
       author = {{{\"U}bler}, Hannah and {Maiolino}, Roberto and {Curtis-Lake}, Emma and {P{\'e}rez-Gonz{\'a}lez}, Pablo G. and {Curti}, Mirko and {Perna}, Michele and {Arribas}, Santiago and {Charlot}, St{\'e}phane and {Marshall}, Madeline A. and {D'Eugenio}, Francesco and {Scholtz}, Jan and {Bunker}, Andrew and {Carniani}, Stefano and {Ferruit}, Pierre and {Jakobsen}, Peter and {Rix}, Hans-Walter and {Rodr{\'\i}guez Del Pino}, Bruno and {Willott}, Chris J. and {Boeker}, Torsten and {Cresci}, Giovanni and {Jones}, Gareth C. and {Kumari}, Nimisha and {Rawle}, Tim},
        title = "{GA-NIFS: A massive black hole in a low-metallicity AGN at z {\ensuremath{\sim}} 5.55 revealed by JWST/NIRSpec IFS}",
      journal = {\aap},
         year = 2023,
        month = sep,
       volume = {677},
          eid = {A145},
        pages = {A145},
          doi = {10.1051/0004-6361/202346137},
archivePrefix = {arXiv},
       eprint = {2302.06647},
 primaryClass = {astro-ph.GA},
       adsurl = {https://ui.adsabs.harvard.edu/abs/2023A&A...677A.145U}
}



\section*{Authors' affiliations}
\noindent
$^{1}$Kavli Institute for Cosmology, University of Cambridge, Madingley Road, Cambridge CB3 0HA, UK
\\
$^{2}$Cavendish Laboratory, University of Cambridge, 19 JJ Thomson Avenue, Cambridge CB3 0HE, UK \\
$^{3}$Department of Physics and Astronomy, University College London, Gower Street, London WC1E 6BT, UK \\
$^{4}$Max-Planck-Institut für extraterrestrische Physik (MPE), Gießenbachstraße 1, 85748 Garching, Germany \\
$^{5}$Centro de Astrobiolog\'ia (CAB), CSIC–INTA, Cra. de Ajalvir km.~4, 28850- Torrej\'on de Ardoz, Madrid, Spain \\
$^{6}$Cosmic Dawn Center (DAWN), Copenhagen, Denmark \\
$^{7}$Niels Bohr Institute, University of Copenhagen, Jagtvej 128, DK-2200, Copenhagen, Denmark \\
$^{8}$Institute of Liberal Arts and Science, Kanazawa University, Kakuma-machi, Kanazawa, 920-1192, Ishikawa, Japan \\
$^{9}$Division of Mathematical and Physical Sciences, Graduate School of Natural Science and Technology, Kanazawa University, Kakuma-machi, Kanazawa, 920-1192, Ishikawa, Japan \\
$^{10}$National Astronomical Observatory of Japan, 2-21-1 Osawa, Mitaka, 181-8588, Tokyo, Japan \\
$^{11}$Dipartimento di Fisica e Astronomia, Università degli Studi di Firenze, Largo E. Fermi 1, 50125, Firenze, Italy \\
$^{12}$INAF — Osservatorio Astrofisico di Arcetri, Largo E. Fermi 5, I-50125, Florence, Italy \\
$^{13}$Dipartimento di Fisica “G. Occhialini,” Università degli Studi di Milano-Bicocca, Piazza della Scienza 3, I-20126
Milano, Italy \\
$^{14}$Department of Astronomy \& Astrophysics, University of California, 1156 High Street, Santa Cruz, CA 95064, USA \\
$^{15}$Center for Astrophysics $|$ Harvard \& Smithsonian, 60 Garden St., Cambridge MA 02138 USA \\
$^{16}$Scuola Normale Superiore, Piazza dei Cavalieri 7, I-56126 Pisa, Italy \\
$^{17}$Waseda Research Institute for Science and Engineering, Faculty of Science and Engineering, Waseda University, 3-4-1, Okubo, Shinjuku, Tokyo 169-8555, Japan \\
$^{18}$Department of Astronomy \& Astrophysics, University of Chicago, 5640 S Ellis Avenue, Chicago, IL 60637, USA \\
$^{19}$Kavli Institute for Cosmological Physics, University of Chicago, Chicago, IL 60637, USA\\
$^{20}$DARK, Niels Bohr Institute, University of Copenhagen, Jagtvej 155A, DK-2200 Copenhagen, Denmark \\
$^{21}$European Space Agency, c/o STScI, 3700 San Martin Drive, Baltimore MD 21218, USA \\
$^{22}$Department of Astronomy, University of Texas, Austin, TX 78712, USA \\
$^{23}$Weinberg Institute for Theoretical Physics, University of Texas, Austin, TX 78712, USA \\ 
$^{24}$Cosmic Frontier Center, The University of Texas at Austin, Austin, TX 78712, USA \\
$^{25}$Department of Physics, University of Oxford, Denys Wilkinson Building, Keble Road, Oxford OX1 3RH, UK \\
$^{26}$Sorbonne Universit\'e, CNRS, UMR 7095, Institut d'Astrophysique de Paris, 98 bis bd Arago, 75014 Paris, France \\
$^{27}$INAF - Osservatorio di Astrofisica e Scienza dello Spazio, Via Piero Gobetti 93/3, 40129, Bologna, Italy\\
$^{28}$Centre for Astrophysics Research, Department of Physics, Astronomy and Mathematics, University of Hertfordshire, Hatfield AL10 9AB, UK \\
$^{29}$Steward Observatory, University of Arizona, 933 N. Cherry Avenue, Tucson, AZ 85721, USA \\
$^{30}$Dipartimento di Fisica, "Sapienza" Universit$\grave{a}$ di Roma, Piazzale Aldo Moro 2, 00185 Roma, Italy \\
$^{31}$INFN, Sezione Roma1, Dipartimento di Fisica, ``Sapienza'' Universit$\grave{a}$ di Roma, Piazzale Aldo Moro 2, 00185, Roma, Italy \\
$^{32}$Department of Astronomy \& Astrophysics, The Pennsylvania State University, University Park, PA 16802, USA \\
$^{33}$AURA for European Space Agency (ESA), ESA Office, Space Telescope Science Institute, 3700 San Matin Drive, Baltimore, MD, 21218, USA \\
$^{34}$Aix Marseille Univ, CNRS, CNES, LAM, Marseille, France \\
$^{35}$INAF - Osservatorio Astronomico di Roma, via Frascati 33, I-00078, Monte Porzio Catone, Italy \\
$^{36}$Space Telescope Science Institute, 3700 San Martin Drive, Baltimore, Maryland 21218, USA \\
$^{37}$Department of Astronomy, University of Geneva, Chemin Pegasi 51, 1290 Versoix, Switzerland \\

\bsp	
\label{lastpage}

\end{document}